\documentclass[prd,nofootinbib,twocolumn,superscriptaddress,floatfix]{revtex4-2}
\usepackage[protrusion=true,expansion=true]{microtype}
\usepackage[T1]{fontenc}
\usepackage{tikzsymbols}
\usepackage{natbib}
\usepackage{float}
\usepackage{rotating} 
\usepackage{graphicx}
\usepackage{color}
\usepackage{mathtools,amssymb,physics}
\usepackage{upgreek}
\usepackage{bm}
\usepackage{mathrsfs}
\usepackage{caption}
\usepackage{pifont}
\usepackage[thinc]{esdiff}
\usepackage{tikz,xcolor}
\usepackage[backref=page]{hyperref}
\hypersetup{
    colorlinks=true,
    citecolor=brown,
    linkcolor=brown,
    filecolor=magenta,      
    urlcolor=blue,
    pdftitle={otoc},
    pdfpagemode=UseThumbs,
}
\definecolor{lime}{HTML}{A6CE39}
\DeclareRobustCommand{\orcidicon}{
	\begin{tikzpicture}
	\draw[lime, fill=lime] (0,0) 
	circle [radius=0.2] 
	node[white] {{\fontfamily{qag}\selectfont \tiny ID}};
	\draw[white, fill=white] (-0.0625,0.095) 
	circle [radius=0.007];
	\end{tikzpicture}
	\hspace{-2mm}
}

\foreach \x in {A, ..., Z}{\expandafter\xdef\csname orcid\x\endcsname{\noexpand\href{https://orcid.org/\csname orcidauthor\x\endcsname}
			{\noexpand\orcidicon}}
}

\newcommand{\be}{\begin{equation}}
\newcommand{\ee}{\end{equation}}
\newcommand{\bea}{\begin{eqnarray}}
\newcommand{\eea}{\end{eqnarray}}

\newcommand{\hrho}{\hat{\rho}}
\newcommand{\trc}{\mathrm{Tr}}

\def\Mp{M_{\rm Pl}}
\def\vk{\textbf{k}}
\def\dx{\mathrm{d}}
\def\wre{w_{\mathrm{re}}}
\def\nre{N_{\mathrm{re}}}
\def\d{\mathrm{d}}

\graphicspath{{./figs/}}

\begin{document}

\title{Primordial cosmic complexity and effects of reheating}

\author{Pankaj Saha\orcidA{}}
\email{saha@seoultech.ac.kr}

\author{Myeonghun Park}
\email{parc.seoultech@seoultech.ac.kr}

\affiliation{School of Natural Sciences, Seoul National University of Science and Technology, \\Seoul 01811, Republic of Korea}

\begin{abstract}
\textit{We study the effects of reheating phase on the evolution of complexities for the primordial curvature perturbation using the squeezed formalism. We examine the evolution of the out-of-time correlator, the quantum discord, and circuit complexity, starting from the inflationary epoch to the radiation-dominated epoch with different reheating scenarios. We find that for a mode that reenters the horizon after reheating, the effect of a finite reheating epoch on the characteristic \textit{freeze-in} amplitude of these primordial complexities can only be distinguished up to three different classes depending on whether the equation of state parameter: $(i)$ $\wre=1/3$ $(ii)$ $\wre<1/3$, or, (iii) $\wre>1/3$. For reheating with different EOS within these classes, the final amplitude will be the same~---~hence the detailed signature of reheating with a class on the complexity measures will be lost. Taking the central value of the scalar spectral index ($n_s=0.9649$) from Planck and the equation of state during reheating $\wre=0.25$ as benchmark values, we found that the behavior of the complexities for all modes smaller than $1.27\times10^{16}\mathrm{Mpc^{-1}}$ can be classified as above. 
However, for the small-scale modes reentering the horizon during reheating, the signature of EOS on the evolution of these two complexities will be embedded in each of the cases separately.}
\end{abstract}

\keywords{Circuit complexity, OTOC}

 \maketitle

\section{\label{intro}Introduction}
The temperature of the cosmic microwave background~(CMB) radiation across the sky is extremely uniform. 
The observed small fluctuations in this temperature map (one part in $10^5$) suggest that the Universe might have been born out of quantum vacuum fluctuations. 
These tiny fluctuations, when enlarged into classical fluctuations on the horizon scale during the colossal expansion during the phase of inflation~\cite{Sato:1980yn,Guth:1980zm,Starobinsky:1980te,Linde:1981mu,Albrecht:1982wi} also provided the seed for large-scale structure formation (see~\cite{baumann2022cosmology} for a review). 
The simplest way to have an inflationary expansion is to consider a scalar field displaced from its minima. As long as the Universe is dominated by the vacuum energy of the scalar field, the Universe expands in a quasi--de Sitter state with weakly broken scale invariance. The inflaton field eventually \textit{rolls down} to the minima ending the inflationary phase and oscillates around the minima of the potential. During this oscillatory phase---known as the reheating---the inflaton decays to other fields and eventually brings in the radiation-dominated hot big bang Universe~\cite{Abbott:1982hn,Dolgov:1982th,Dolgov:1989us,Traschen:1990sw,Kofman:1994rk,Shtanov:1994ce}.
\par 
The observed patterns in the CMB~\cite{dodelson2020modern,durrer2020CMB} are classical, and the seed primordial perturbations can be either quantum or classical in nature~\cite{Lim:2014uea,Maldacena:2015bha,Martin:2015qta,Green:2020whw}. As a matter of fact, the classical stochastic fields match all statistics if we only account for the Gaussianity. However, as pointed out in~\cite{Green:2020whw}, a detection of primordial non-Gaussianity can confirm the quantum origin of the cosmic structure (for recent developments in this context, see, for instance,~Refs. \cite{Agullo:2022ttg,Green:2022fwg,Ghosh:2022cny,Micheli:2022tld,Colas:2022kfu} for some interesting developments). Assuming that the primordial fluctuations originated from quantum vacuum fluctuations, their transition to classical perturbations is studied in the parlance of \textit{squeezing formalism} where the perturbations got \textit{squeezed} as they exit the horizon~\cite{Polarski:1995jg, Albrecht:1992kf}. After squeezing, the mode functions can be treated as real---consequently, the commutator vanishes. The squeezing results in the dominance of the growing modes of the perturbations over the decaying modes, implying that inflation generates particles out of the initial Bunch-Davies (BD) vacuum. Further classicalization happens when all modes leave the horizon at the end of inflation. All matter (including dark matter), radiation, and the Standard Model~(SM) degrees of freedom must, in principle, be created during reheating. The memory of the initial quantum nature of these perturbations is transferred to classical perturbation during thermalization. 
\par
\begin{figure*}[ht!]
    \centering
    \vspace{0.2cm}
    \includegraphics[scale=1.6]{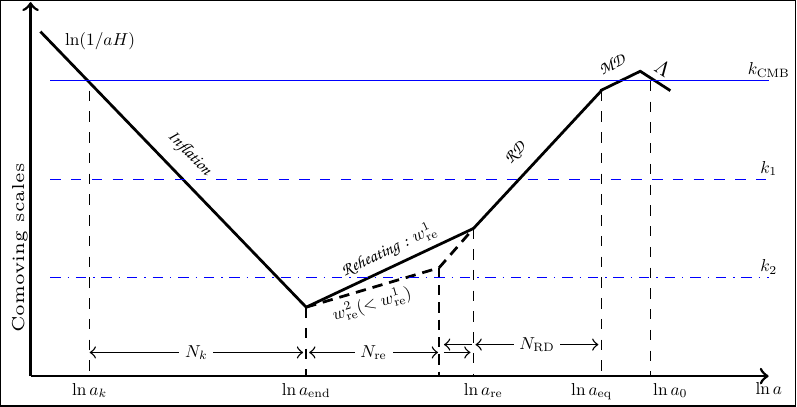}
    \caption{We show the evolution of the comoving Hubble horizon for two different reheating scenarios. RD, MD, and $\Lambda$ are, respectively, the radiation, matter, and the recent dark-dominated phases. The $k_{\rm CMB}$ is the pivot scale at which Planck determines $n_s$. $k_1$ and $k_2$ are two modes that rerenter the horizon during RD and reheating respectively. As fixing $n_s$ corresponds to picking a $N_k$ and the reheating history, then changing $\wre$ will result in changing the $e$-folds during reheating. Thus, for the mode that reenters the horizon during RD, such as $k_1$, changing the reheating history will not affect when the mode reenters the horizon. As the squeezing grows only when the modes are outside the horizon, hence, for such modes, the change in the equation of state (EOS) within the class of reheating will not have any effect on squeezing, and consequently, all the co-relation functions will be identical.}
    \label{fig:illustation_reentry}
\end{figure*}
The reheating phase is episodic: the first stage is dominated by the coherently oscillating inflaton condensate. The product of the daughter fields is mainly due to the parametric resonance. The second stage initiates when the produced fields start to backreact significantly; the inhomogeneities grow and contribute to a sizeable amount of total energy. The final phase is the stationary phase when the decay of inflaton via nonperturbative processes becomes inefficient. Now, for a complete decay of the inflaton and to bring in the radiation-dominated (RD) Universe, a perturbative decay channel became necessary for many scenarios~\cite{Abbott:1982hn,Dolgov:1982th,Kofman:1994rk,Podolsky:2005bw,Maity:2018qhi}. Nevertheless, a macroscopic description of the reheating phase can be prudently described by tracking the dilution of the energy with the expansion of the Universe~\cite{Lorenz:2007ze,Martin:2010kz,Adshead:2010mc,Mielczarek:2010ag,Easther:2011yq,Martin:2014nya,Dai:2014jja,Drewes:2017fmn}. This can be uniquely specified by knowing the equation of the state parameter~(EOS) for the effective fluid describing the energy density of the Universe. For a fluid with EOS $w$, the energy density dilutes as $\rho\propto a^{3/2(1+w)}$. The equation of state for the initial phase of the reheating phase is determined by the dominant term of the potential during reheating. For an inflaton condensate oscillating coherently in a potential $V(\phi)\propto\phi^n$, the effective equation of the state of the system is $w_{\rm re}=(n-2)/(n+2)$~\cite{Mukhanov:2005sc,Saha:2020bis}. Numerical studies show that for quadratic potential, the EOS increases from this coherent oscillating EOS ($w_{\rm co}$) value of zero and reaches a maximum of around $0.25$ and finally again transits to zero with long-time simulation~\cite{Podolsky:2005bw,Maity:2018qhi,Saha:2020bis,Antusch:2020iyq}. For other models ($n>2$), the general behavior is that the value of EOS---starting from $w_{\rm co}$---reaches a radiation-like EOS~\cite{Lozanov:2016hid,Maity:2018qhi,Saha:2020bis,Antusch:2020iyq}. Accordingly, the EOS during reheating is determined by the shape of the inflationary potential; in principle, other values of reheating EOS are also possible due to interaction between different fields. In this work, as a most conservative bound, we will take the values of reheating EOS between $(0,1)$. At the end of reheating, the equation of state reaches radiation-like behavior and formally starts the radiation-dominated hot Big Bang evolution. At the background level, the reheating epoch can be described by the dependence of the Hubble expansion on the equation of state parameter. The crucial point in our study can now easily be understood with the help of the illustration in Fig. \ref{fig:illustation_reentry}. The perturbation modes evolve differently based on whether they are inside or outside of the horizon. Thus, the duration of reheating determines when a mode will reenter the horizon. However, by consistently taking the reheating constraints into account, we find that for modes that reenter the horizon at the radiation-dominated era (this includes the large-scale modes that we observe in CMB anisotropy which is roughly $10^{-5}~\mathrm{Mpc^{-1}}<k<1~\mathrm{Mpc^{-1}}$), the reentry is not affected by the individual values of the EOS; rather, it depends on whether (i) $\wre<1/3$, (ii) $\wre=1/3$ or, (iii) $\wre>1/3$. Consequently, the modes reentering the horizon during the RD and the reheating effects are classified into three classes.
\par
In this work, based on the above observation, we study the effects of the reheating phase on the evolution of various measurements of complexities for primordial curvature perturbation. We analyze the Squeezed out-of-time-order correlator~(OTOC) that has been explored extensively in regards to studying the quantum aspect of chaos or scrambling~\cite{Maldacena:2015waa,Rozenbaum:2017zfo,Hashimoto:2017oit,Gharibyan:2018fax,Rozenbaum:2019kdl}. We also examine the quantum discord for the bipartite system $\mathcal{E}=\mathcal{E}_{\vk}\bigotimes\mathcal{E}_{-\vk}$ [i.e., the two-mode ($\vk,-\vk$) pairs of the curvature perturbation]. The study of quantum discord has recently gained a lot of interest in the Early Universe cosmology as a means of exploring the quantumness of various primordial correlation functions~\cite{Lim:2014uea,Martin:2015qta}. For the current system, the quantum discord is identical to the quantum entanglement or von Neumann entropy of the bipartite system. A third measure of primordial complexity we consider is the circuit complexity---the rate of change of which, with respect to cosmic time, can be identified as the quantum Lyapunov exponent. Thus, it can be directly used to study the development of quantum chaos in a system. Now, as the evolution of these complexity measures follows the evolution of the squeezing parameters, we observe that both complexity measures follow the same patterns of evolution in the early Universe.
Comprehensive studies of different complexity measures and their theoretical aspects in the context of cosmology can be found in ~\cite{Martin:2019wta,Bhattacharyya:2020rpy,Bhattacharyya:2020kgu,Choudhury:2020yaa,Haque:2020pmp,Lehners:2020pem,Choudhury:2020hil,Adhikari:2022oxr}.
\par 
We have structured the remainder of the paper as follows: In Sec~\ref{sec:background}, we will describe the background inflationary model. The details of the reheating constraints have been presented in Sec~\ref{sec:reh_cons}. We then develop the squeezing formalism for the primordial curvature perturbation in Sec~\ref{sec:sqz_prim}. Various complexity measures and their relationship with the squeezing parameters are described in Sec~\ref{sec:pcm}. We conclude in Sec~\ref{sec:res_dis}. We will consider $\hbar = c = 1$ unless otherwise stated. We have used $\Mp( = 1/\sqrt{8\pi G}=2.43\times10^{18}~\mathrm{GeV})$ to denote the reduced Planck mass. We will take the usual Friedmann-Lema\^{i}tre-Roberson-Walker (FLRW) metric as our background metric $\dd s^2= \dd t^2 -a^2(t)(\dd x^2+\dd y^2+\dd z^2) $ for deriving our equations, where $a(t)$ is the scale factor and $t$ represents the cosmic time.

\section{The background Evolution}\label{sec:background}
We will use the ubiquitous $m^2\phi^2$ model to describe the background evolution. The choice of an inflationary model is not very important here as the general conclusions will hold true for any canonical single-field models. Additionally, the model has simple analytic solutions of the background that will greatly expedite the later numerical solutions of the differential equations for the squeezing parameters. We will consider the following scenario: 
\begin{itemize}
    \item Inflation is due to the slow-rolling scalar field described by $m^2\phi^2$ type potential. 
    \item The reheating phase is described by an effective fluid with an equation of the state parameter $\wre$.
    \item After reheating, the Universe will transit to the radiation-dominated phase. We will use the techniques developed in~\cite{Dai:2014jja} to find the inflationary($N_k$) and reheating $e$-folds ($N_{\rm re}$). We will see that taking the reheating $e$-folds consistently is crucial in finding out when a mode reenters the horizon after inflation. 
\end{itemize}
\par
The action for the scalar field is given by
\begin{equation}
  S = \int \sqrt{-\abs{g}}dx^4 \bigg[\frac{\Mp^2}{2}R + \frac{1}{2}\partial_{\mu}\phi\partial^{\mu}\phi+ V(\phi)\bigg]
  \label{eq:action}
\end{equation}
During inflation, the energy is dominated by the vacuum energy of the inflaton. We vary the action in (\ref{eq:action}) for the metric and the scalar field $\phi$ to arrive at the following background equations:
\begin{align}
    3\Mp^2H^2 \equiv \left(\frac{\dot{a}}{a}\right)^2 &= \left[\frac{1}{2}\dot{\phi}^2 + V(\phi)\right],
    \label{eq:fr}\\
    \ddot{\phi} + 3H\dot{\phi} + V'(\phi) &= 0.
    \label{eq:kg}
\end{align}
The condition for inflation is quantified by defining the so-called slow-roll parameters, viz:
\begin{align}
\epsilon_1 &= \frac{\d \ln H}{\d N}, \\
\epsilon_{i+1} &= \frac{\d \ln \epsilon_i}{\d N}~~~\text{for $i > 1$},
\end{align}
where during inflation, these slow-roll parameters remain small $\abs{\epsilon}\ll1$ while the end of inflation is marked by $\epsilon_1=1$.
\par
For the potential considered here, we can write the analytic solutions of Eqs. (\ref{eq:fr} and \ref{eq:kg})~as~\cite{Mukhanov:2005sc,Peter:2013avv}
\begin{align}
 H = \begin{cases}
        m_{\phi}\sqrt{\frac{1}{3} - \frac{2}{3}\ln a}, & a\leq1\\
        \frac{m_{\phi}}{\sqrt{3}}\mathrm{e}^{-\frac{3}{2}(\wre+1)\ln a},& 1\leq a\leq a_{\rm re}\\
        \frac{m_{\phi}}{\sqrt{3}}\mathrm{e}^{-\frac{\ln a_{\rm re}}{2}(3\wre-1)}\mathrm{e}^{-2\ln a},  & a\geq a_{\rm re}
        \end{cases}
        \label{eq:hubble}
\end{align}
\begin{figure}
    \centering
    \includegraphics[scale=0.75]{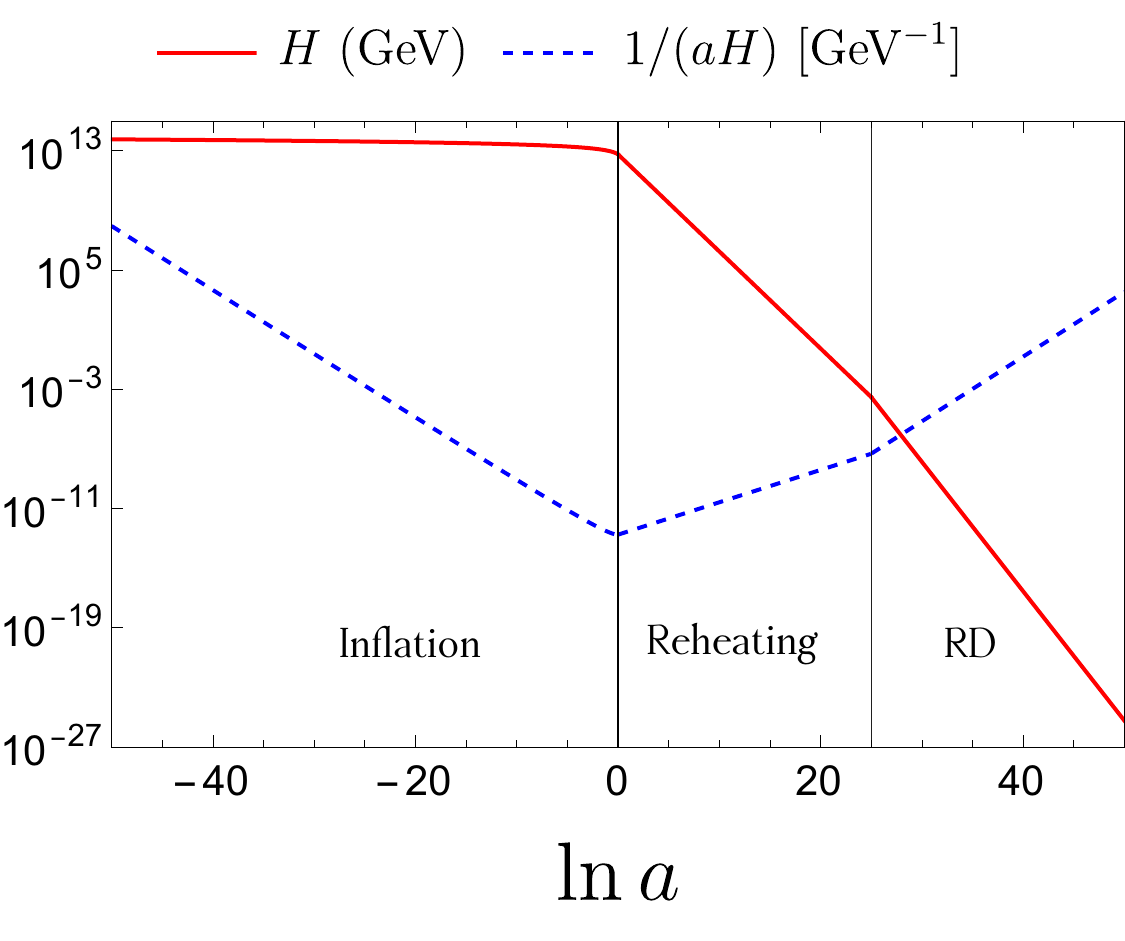}
    \caption{Evolution of the Hubble scale and the comoving horizon as described in Eq.~(\ref{eq:hubble}) during the phase of inflation, reheating described by $\wre=0$ and the radiation-dominated phase.}
    \label{fig:scales}
\end{figure}
where we normalized the scale factor such that the scale factor at the end of inflation is $a_{\rm E}=1$. The evolution of the Hubble parameter and the comoving Hubble horizon for the model is shown in Fig. \ref{fig:scales}.

We will take the EOS during reheating as a free parameter while the reheating $e$-folds will be calculated $\acute{a}$ $la$ reheating constraints in the same spirit as~\cite{Lorenz:2007ze,Martin:2010kz,Adshead:2010mc,Mielczarek:2010ag,Easther:2011yq,Martin:2014nya}, particularly using the techniques developed in~\cite{Dai:2014jja}. We will elaborate in the next section that the reheating constraints play a critical role in deciding the fate of primordial correlations by determining when the perturbation modes reenter the horizon.

\section{\label{sec:reh_cons}Reheating Constraints}
The expansion history during reheating is largely uncertain. In order to track the expansion history of the Universe during this phase, we parametrize the effective equation of state from the end of inflation to the time when the Universe thermalized completely (i.e., the end of reheating) as 
\[\wre = \frac{1}{\nre}\int_0^{\nre} w(N)dN\]
The equation of state during reheating must be larger than $-1/3$ so as to end the inflationary expansion, and it is assumed to be smaller than 1 in order to obey causality. 
An equation of state larger than $1/3$ is challenging to come up with as it requires a potential dominated by high-dimensional potentials (i.e., potentials of the form or higher than $\phi^6$) near the bottom of the potential---unnatural from a quantum field theory perspective. 
\begin{figure}[t!]
    \centering
    \includegraphics[scale=0.7]{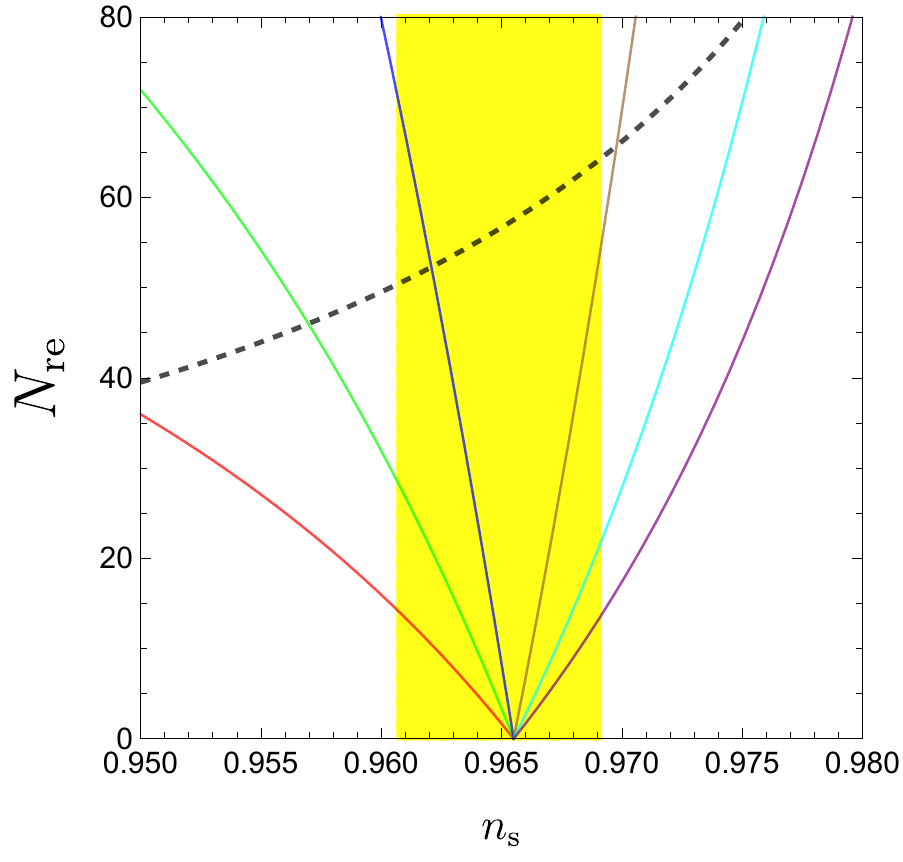}
    \caption{Variation of $\nre$ with the scalar spectral index for different EOS parameters: $\wre=-1/3$~(red), $\wre=0$~(green), $\wre=0.2$~(blue), $\wre=0.5$~(brown), $\wre=0.75$~(cyan) and $\wre=1$~(purple). The black dashed line is the $N_k$. We show the Planck 1-$sigma$ level with the yellow shaded region~\cite{Planck:2018jri}.}
    \label{fig:reh_cons}
\end{figure}
Still, as a conservative range, we can consider the EOS parameter to be within the range of $-1/3$ and 1.
Given the inflationary model and the $\wre$, the essential information we will require for our later analysis is the number of inflationary $e$-folds $N_k$ and the reheating $e$-folds $\nre$.
To this end, we notice, following~\cite{Dai:2014jja,Cook:2015vqa}, that the perturbation modes we observe today are those of comparable scale to the present horizon size. Planck, for instance, determines the scalar spectral index $n_s$ at the pivot scale $k=0.05~{\rm Mpc}^{-1}$. 
Consequently, we relate the comoving Hubble scale $a_kH_k = k$ when a mode $k$ (in this case, we are concerned with the pivot scale) exits the horizon during inflation to that of the present time as:
\begin{align}
    \frac{k}{a_0 H_0}=\frac{a_k}{a_{\rm end}}\frac{a_{\rm end}}{a_{\rm re}}\frac{a_{\rm re}}{a_{\rm eq}}\frac{a_{\rm eq}H_{\rm eq}}{a_0 H_0}\frac{H_k}{H_0}.
    \label{eq:kaH}
\end{align}
where the quantities with subscript $k$ referred to the time of horizon exit and other subscripts are for the following instances: the end of inflation ($\rm end$), end of the reheating (or beginning of the radiation-dominated era: $\rm re$), radiation-matter equality ($\rm eq$), and the present time $0$. 
We define the $e$-folds as between two instances when the scale factor grows from $a_{\rm i}$ to $a_{\rm f}$ as $N_{\rm f}=\ln(a_{\rm f}/a_{\rm i})$.

Now for the model, the number of inflationary $e$-folds when the pivot scale $k$ exit the horizon is given by~\cite{Cook:2015vqa}:
\begin{equation}
    N_k = \frac{4}{2(1-n_s)}
    \label{eq:nk}
\end{equation}
while the reheating $e$-folds can be found in terms of the model parameters and the EOS parameter during reheating~\cite{Dai:2014jja}:
\begin{figure}[t!]
    \centering
    \vspace{0.2cm}
    \includegraphics[scale=0.9]{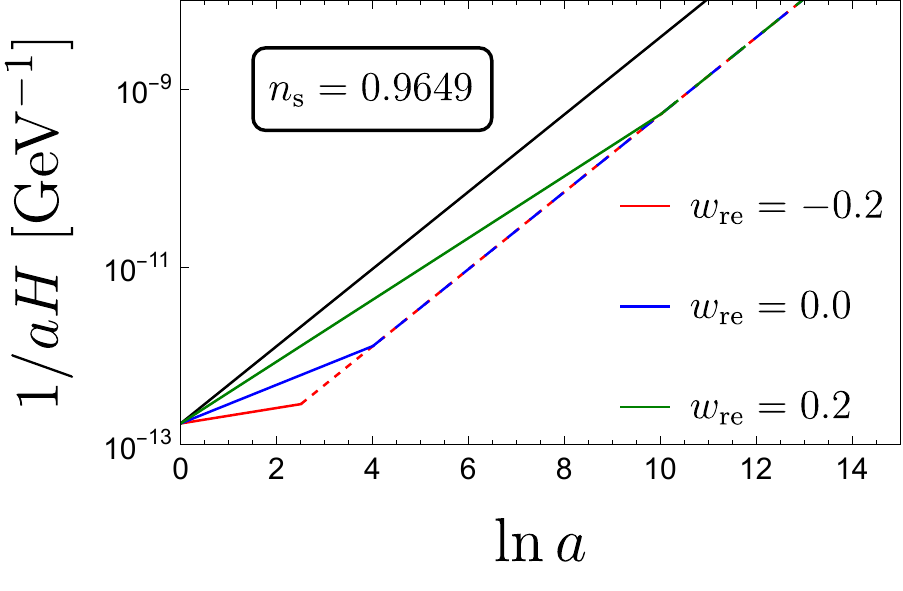}
    \caption{We plot the evolution of the comoving Hubble horizon from the end of inflation for three different EOS parameters. We fix the $n_s$ to be the central value from Planck measurement~\cite{Planck:2018jri} and find the reheating $e$-folds $\nre$ using~(\ref{eq:nre}).}
    \label{fig:horizon_post}
\end{figure}
\begin{align}
\nonumber
    N_{\rm re} =& \frac{4}{3\wre-1}\Bigg[N_k + \ln(\frac{k}{a_0T_0}) + \frac{1}{4}\ln(\frac{40}{\pi^2g_{\ast}})\\
    &+ \frac{1}{3}\ln(\frac{11g_{s,\ast}}{43}) - \frac{1}{2}\ln(\frac{\pi^2\Mp^2 r A_s}{2V_{\rm end}^{1/2}})\Bigg]\\
    =& \frac{4}{3\wre-1}\Bigg[N_k - 61.6 - \frac{1}{2}\ln(\frac{\pi^2\Mp^2 r A_s}{2V_{\rm end}^{1/2}})\Bigg]
    \label{eq:nre}
\end{align}
where we took, $g_{\rm re} = 100$ and used the Planck pivot of $k=0.05~{\rm Mpc^{-1}}$. In the above relation, all the values are known, and for the model at hand, we can express the tensor-to-scalar ratio:
\begin{equation}
    r_k = 4(1-n_s)
    \label{eq:rk}
\end{equation}
Thus, we know the inflationary $e$-folds and the reheating $e$-folds once the spectral tilt $n_s$ is known. 
\par
We plot the variation $\nre$ for different values of reheating EOS parameters for the model in Fig. \ref{fig:reh_cons}. The variation of $N_k$ with $n_s$ is also shown in the same plot as the black dashed line. While taking the central value of $n_s$ from Planck as a benchmark, the evolution of the comoving Hubble horizon is shown in Fig.\ref{fig:horizon_post}. The qualitative nature of these plots is essentially the same for all canonical inflationary models. Consequently, using these plots, we can deduce some general features of the reheating phase that will be useful for our later discussion. (i) Once we fix the spectral tilt, the reheating EOS can be either $\wre<1/3$ or $\wre>1/3$. Also, for a specific value of $n_s$, all these lines merge at $\nre=0$, i.e., the instantaneous reheating. This special case of instantaneous reheating can more intuitively be thought of as reheating with $\wre=1/3$. (ii) Knowing $n_s$ (consequently $\nre$), we also see that the postinflationary evolution of the Hubble horizon is uniquely determined within the degeneracy of the reheating EOS in the subclass. However, after reheating, the evolution of the horizon is unique. To consider a realistic scenario, we can take $\wre=0.25$ (the maximum value of the EOS from lattice simulation results for preheating with potentials which behave as $\phi^2$ around the minima~\cite{Podolsky:2005bw,Maity:2018qhi,Saha:2020bis}). In such a case, all modes with $k<1.27\times 10^{16}~\mathrm{Mpc^{-1}}$ will reenter the horizon during radiation-dominated era. A general model-independent argument of the dependence of reheating EOS on the horizon reentry of various perturbation modes is provided in Fig. \ref{fig:illustation_reentry}. With a description of the background inflationary model and the reheating constraints, we are now in a position to describe the squeezing formalism of cosmological perturbation, which will be done in the next section.

\section{\label{sec:sqz_prim}Squeezing the primordial perturbations}
The squeezing formalism was first used in quantum optics~\cite{Walls:1983zz,Schumaker:1986bl} to describe the phenomenon when a specific phase space volume evolves in a way such that it contracts in one direction while simultaneously expanding in another direction. 
Authors in~\cite{Albrecht:1992kf} applied the notion of squeezing to inflationary perturbations (see also~\cite{Grishchuk:1989ss,Grishchuk:1990bj,Polarski:1995jg,Mijic:1998if,Haro:2008zz} in the context). To make ourselves familiar with the terminology, we will first present a brief review of the description of cosmological perturbations as two-mode squeezed states following~\cite{Polarski:1995jg} (see also~\cite{Martin:2004um,Martin:2007bw} for review). 
We take the background metric as the spatially flat FLRW metric,
\begin{equation}
 \label{eq:metric0}
 ds^2 = -dt^2 + a(t)^2d\vec{x}^2 = a(\eta)^2\left(-d\eta^2 + d\vec{x}^2\right),
\end{equation}
where $\eta$ is the conformal time defined through the relation $d\eta=dt/a(t)$. We will consider the matter sector consisting of a single scalar field with the canonical kinetic term. We decompose the field into a homogeneous time-dependent background field and small perturbations $\phi(\vec{x},t) = \phi_0(t) + \delta\phi(\vec{x},t)$. The fluctuations in the metric are written likewise as:
\begin{equation}
 \label{eq:metric}
 ds^2 = a(\eta)^2\left[-(1+2\Psi(x,\eta))d\eta^2 + (1-2\Psi(x,\eta))d\vec{x}^2\right].
\end{equation}
where $\Psi$ is the scalar degree of freedom of the metric.
Now writing the gauge-invariant combination known as the curvature perturbation $\mathcal{R} = \Psi + \frac{H}{\dot{\phi}_0}\delta\phi$, the action for the scalar perturbation takes the following simple form:
\begin{equation}
 \label{eq:actionR}
S = \frac{\Mp^2}{2}\int \dx t\dx^3x a^3\frac{\dot{\phi}_0}{H^2}\left[\dot{\mathcal{R}}^2 - \frac{1}{a^2}(\partial_i\mathcal{R})^2\right].
\end{equation}
Using the Mukhanov-Sasaki variable\footnote{The study of tensor perturbations can be done in an exactly similar way with $z'/z$ replaced by $a'/a$.} $v\equiv z\mathcal{R}$, with $z=\Mp a\sqrt{2\epsilon_1}$. Using the conformal time, the above action reduces to the action of a harmonic oscillator with time-dependent mass:
\begin{equation}
\label{eq:actionv}
 S = \frac{1}{2}\int \dx\eta \dx^3x\left[v'^2 - (\partial_iv)^2+\left(\frac{z'}{z}\right)v^2 -2\frac{z'}{z}v'v\right]
\end{equation}
where the prime denotes derivatives with respect to conformal time. Using the canonical momentum $\pi(\eta)=v'^2 - \frac{z'}{z}v$, we construct the Hamiltonian (in Fourier space) as
\begin{equation}
 \label{eq:hamiltonianR}
 \mathscr{H} = \int_{\mathbb{R}^{3+}} \dx^3\vk\left[\pi_{\vk}\pi_{\vk}^{\ast} + k^2v_{\vk}v_{\vk}^{\ast} + \frac{z'}{z}(v_{\vk}\pi_{\vk}^{\ast} + \pi_kv_{\vk}^{\ast})\right].
\end{equation}
One important aspect of the Fourier modes is that since $v(\eta,\mathbf{x})$ is real, one must have $v_{-\vk} = v^{\ast}_{\vk}$, implying only half of the modes in Fourier space are independent. Thus, with the Hamiltonian defined in (\ref{eq:hamiltonianR}), we have summed up only in half the Fourier space, i.e., $\vk\in\mathbb{R}^{3+}$~\cite{cohen-tannoudji1989,Martin:2015qta}.
The equations describing the time evolution of the coordinate and momentum are 
\begin{align}
 \label{eq:eqvk1}
 v_k' &= \frac{\partial\mathscr{H}}{\partial\pi_k} = \pi + \frac{z'}{z}v_k\\
 \label{eq:eqvk2}
 \pi_k' &= -\frac{\partial\mathscr{H}}{\partial v_k} = -k^2v_k - \frac{z'}{z}\pi_k.
\end{align}
Now we quantize the fields and define the field and the conjugate momentum in terms of the so-called ladder operators (employing the Heisenberg picture):
\begin{align}
 \label{eq:lagger1}
 \hat{v}_{\vk}(\eta) &= \frac{1}{\sqrt{2k}}\left(\hat{a}_{\vk}(\eta) + \hat{a}^{\dagger}_{-\vk}(\eta)\right)\\
 \label{eq:lagger2}
 \hat{\pi}_{\vk}(\eta) &= -i\sqrt{\frac{k}{2}}\left(\hat{a}_{\vk}(\eta) - \hat{a}^{\dagger}_{-\vk}(\eta)\right),
\end{align}
where the ladder operators inherit their usual commutator relations: $[\hat{a}_{\vk},\hat{a}^{\dagger}_{\vk'}] = \delta^{(3)}(\vk - \vk')$ from the commutators of the field and its conjugate momentum. To find the time evolution of the ladder operators, we insert Eqs. (\ref{eq:lagger1}) and \ref{eq:lagger2}) into Eqs.(\ref{eq:eqvk1}) and (\ref{eq:eqvk2}). This yields
\begin{align}
 \label{eq:eomladder1}
 \hat{a}_{\vk}'(\eta) &= -i k\hat{a}_{\vk}(\eta) + \frac{z'}{z}\hat{a}^{\dagger}_{-\vk}(\eta),\\
 \label{eq:eomladder2}
 \hat{a}^{\dagger\prime}_{-\vk}(\eta) &= \frac{z'}{z}\hat{a}_{\vk}(\eta) + i k\hat{a}^{\dagger}_{-\vk}(\eta)
\end{align}
To solve these coupled differential equations, we take the Bogolyubov (also transliterated as ``Bogoliubov'') transformation of the initial time ladder operators to write them in any instant of time as
\begin{align}
 \label{eq:bt1}
 \hat{a}_{\vk}(\eta) &= \alpha_k(\eta)\hat{a}_{\vk}(\eta_0) + \beta_k(\eta)\hat{a}^{\dagger}_{-\vk}(\eta_0),\\
 \label{eq:bl2}
 \hat{a}^{\dagger}_{-\vk}(\eta) &= \alpha^{\ast}_k(\eta)\hat{a}^{\dagger}_{-\vk}(\eta_0) + \beta^{\ast}_k(\eta)\hat{a}_{\vk}(\eta_0),
\end{align}
Also, we note that as the commutation relations for the ladder operators shall hold at all times, we must have $|\alpha_k({\eta})|^2 -|\beta_{k}(\eta)|^2=1$. Considering this, we can parametrize the coefficient functions $\alpha_k(\eta)$ and $\beta_k(\eta)$ with the help of the so-called \textit{squeezing parameters}---$r_k$, $\varphi_k$ and $\theta_k$---as:
\begin{align}
 \label{eq:paramak1}
 \alpha_k(\eta) &= e^{{-i\theta_k(\eta)}}\cosh(r_k(\eta)),\\
 \label{eq:paramak2}
 \beta_k(\eta) &= -e^{ i\left(\theta_k(\eta) + 2\varphi_k(\eta)\right)}\sinh(r_k(\eta)),
\end{align}
Next, we want to determine the time-evolution equations for these squeezed parameters. For this we first substitute the Bogolyubov transformation for $\alpha_k$ and $\beta_k$ as in (\ref{eq:bt1}-\ref{eq:bl2}) into the time-evolution equations in Eqs.(\ref{eq:eomladder1}) and (\ref{eq:eomladder2}) to get
\begin{align}
 \label{eq:eqak1}
 \alpha_k' &= \frac{z'}{z}\beta^{\ast}_k - ik\alpha_k,\\
 \label{eq:eqak2}
 \beta_k' &= \frac{z'}{z}\alpha^{\ast}_k - ik\beta_k,
\end{align}
while taking the derivatives on Eqs.(\ref{eq:paramak1}) and (\ref{eq:paramak2}) yields
\begin{align}
 \label{eq:eqak11}
 \alpha_k' &= e^{-i\theta_k}\left(r_k'\sinh r_k -i\theta_k'\cosh r_k\right),\\
 \label{eq:eqak21}
 \beta_k' &= -e^{i(\theta_k + 2\varphi_k)}\left(r_k'\cosh r_k +i(\theta_k' + 2\varphi_k')\sinh r_k\right),
\end{align}
Now, these identities, when combined in appropriate combinations such as 
$$\frac{1}{2}\left(e^{i\theta_k}\alpha_k' + e^{-i\theta_k}\alpha_k'^{\ast}\right),$$ 
$$\frac{1}{2}\left(e^{i\theta_k}\alpha_k' - e^{-i\theta_k}\alpha_k'^{\ast}\right),$$
and 
$$\frac{1}{2}\left(e^{-i(\theta_k + 2\varphi_k)}\beta_k' - e^{i(\theta_k + 2\varphi_k)}\beta_k'^{\ast}\right)$$ fetch us the time-evolution equations for the three squeezing parameters:
\begin{align}
\label{eq:rk}
 r_k'       &= -\frac{z'}{z}\cos 2\varphi_k,\\
 \label{eq:phik}
 \varphi_k' &= -k + \frac{z'}{z}\coth 2r_k\sin 2\varphi_k,\\
 \label{eq:thetak}
 \theta_k' &= k - \frac{z'}{z}\tanh r_k\sin 2\varphi_k
\end{align}
Now with the mode $f_k(\eta) = \frac{1}{\sqrt{2k}}(\alpha_k(\eta) + \beta_k^{\ast}(\eta))$, we can write the field operator as
\begin{align}
\nonumber
 \hat{v}(\eta) &= f_k(\eta)\hat{a}_k(\eta_0) + f^{\ast}_k(\eta)\hat{a}^{\dagger}_{-k}(\eta_0),\\
  \label{eq:field_op}
               &= \sqrt{2k}\mathrm{Re}\left(f_k(\eta)\right)\hat{v}_k(\eta_0) + \sqrt{\frac{2}{k}}\mathrm{Im}\left(f_k(\eta)\right)\hat{\pi}_k(\eta_0) 
\end{align}
To get the initial conditions for the squeezing parameter $r_k$, $\varphi_k$ and $\theta_k$, we compare the mode function of perturbation,
\begin{align}
\nonumber
 f_k(\eta) &= \frac{1}{\sqrt{2k}}\left[\alpha_k(\eta) + \beta^{\ast}_k(\eta)\right],\\
 \nonumber
 &= \frac{1}{\sqrt{2k}}\Bigg[{\rm e}^{-i\theta_k(\eta)}\cosh(r_k(\eta))\\ 
 &- {\rm e}^{-i(\theta_k(\eta)+2\varphi_k(\eta))}\sinh(r_k(\eta))\Bigg]
\end{align}
to the Bunch-Davies vacuum:
\begin{equation}
 \label{eq:bdvacuum}
 u_k(\eta) = \frac{{\rm e}^{-ik\eta}}{\sqrt{2k}}\left(1-\frac{i}{k\eta}\right).
\end{equation}
This fetches us with the following form of the squeezing parameters that coincide with the Bunch-Davies vaccum~\cite{Kuss:2021gig}:
\begin{align}
 \label{eq:bdsqz1}
 r_k &= -{\rm arcsinh}\left(\frac{1}{2k\eta}\right),\\
  \label{eq:bdsqz2}
 \varphi_k &= \frac{\pi}{4} - \frac{1}{2}{\rm arctan}\left(\frac{1}{2k\eta}\right),\\
  \label{eq:bdsqz3}
 \theta_k &= k\eta + {\rm arctan}\left(\frac{1}{2k\eta}\right),
\end{align}

Alternatively, using the squeezing parameters, we can also define the familiar two-mode squeeze ($\hat{\mathcal{S}}_{\vk}$) and rotation ($\hat{\mathcal{R}}_{\vk}$) operators as
\begin{align}
    \label{eq:sqz_op}
    \mathcal{\hat{S}_{\vk}}(r_k, \phi_k) &\equiv \exp\left[\frac{1}{2}\left(z_k^{\ast}\hat{a}_{\vk}\hat{a}_{-\vk}-z_k\hat{a}^{\dagger}_{-\vk}\hat{a}^{\dagger}_{\vk}\right)\right],\\
    \mathcal{\hat{R}_{\vk}}(\theta_k) &\equiv \exp\left[-i\theta_k(\eta)\left(\hat{a}_{\vk}\hat{a}^{\dagger}_{\vk} + \hat{a}^{\dagger}_{-\vk}\hat{a}_{-\vk}\right)\right]
    \label{eq:rot_op}
\end{align}
where $z_k = r_k(\eta)e^{i\phi_k(\eta)}$.
This, eventually, can be combined to define the unitary time evolution operator:
\begin{equation}
    \hat{\mathcal{U}}_{\vk} = \hat{\mathcal{S}}_{\vk}(r_k, \phi_k)\hat{\mathcal{R}}_{\vk}(\theta_k).
\end{equation}
Now, with the definition of squeezing parameters and their evolution equations at hand, we will define various complexity measures and express them in terms of the squeezing parameters in the section that follows.

\section{\label{sec:pcm}Primordial Complexity Measures}
Complexity is a prevalent topic at the junction of many topics, such as theoretical computer science, quantum computing, information processing, many-body quantum theory, and black hole physics. By and large, complexity quantifies the resources required to implement a computation. In theoretical computer science, complexity classifies the Boolean functions according to the size or depth of the Boolean circuits employed to compute them~\cite{Arora:2009ccx,Jukna:2012bcx,Vollmer:2013icc}. The circuit model in quantum computing provides the natural measure of quantum circuit complexity of pure states and unitary transformations. A pure state can be defined as the size of the smallest circuit that produces the state from a product state. Similarly, for a unitary transformation, the complexity can be measured with the number of gates of the smallest circuit that affects the unitary.
There are different methods for measuring the complexity of a system. For a pure state, these criteria may be equivalent (for a recent study comparing different methods, see~\cite{Martin:2022kph}). Below, we will study the complexity of the primordial fluctuations employing two different methods, namely, using the out-of-time-order correlator and quantum discord.
\subsection{Out-of-time-order correlator as a diagnostics of quantum chaos}
The high sensitivity of initial conditions characterizes chaos in classical systems. In a chaotic system, two nearby trajectories with small perturbations in the initial conditions diverge exponentially. The Poisson bracket between the position and momentum variables is a go-to tool to capture such behavior,
\begin{equation}
 \{q(t), p(0)\}^2 = \left(\frac{\partial q(t)}{\partial q(0)}\right)^2 \sim \sum_{i} c_n e^{2\lambda_i t},
 \label{eq:poisson}
\end{equation}
where $\lambda_i$ is the Lyapunov characteristic exponents of the system. For quantum systems, quantifying the chaos may be a bit challenging. The simple analog of the Poisson
bracket defined in (\ref{eq:poisson}) for quantum systems is the {\it unequal time commutator} $[\hat{q}(t), \hat{p}(0)]$ that reduces to the Poisson bracket ($\sim i\hbar\{q(t), p(0)\}$) in the semiclassical limit. Thus, quantum chaos could be studied from this quantity. However, being an operator and not a $c$-number, a more useful quantity to study the chaos in quantum systems is the double unequal-time commutator or the OTOC
\begin{equation}
 \mathcal{C}^{T}(t) \equiv -\langle[\hat{q}(t),\hat{p}(0)]^2\rangle_{\beta},
 \label{eq:otoc0}
\end{equation}
where the average over some state $\langle~\rangle_{\beta}$ is usually taken as a thermal average at temperature $T=1/\beta$. This is generalized to define the OTOC of any two Hermitian operators as
\begin{equation}
 \mathcal{C}^{T}_{~A,B} = -\langle[\hat{A}(t),\hat{B}(t)]^2\rangle_{\beta}
 \label{eq:otoc1}
\end{equation}
At this point, we must also note that, for the cases considered here, as the commutators are just $c$-numbers implying the results will be independent of the averaging process. Additionally, we will calculate the OTOC for the general class of states known as the {\it squeezed quantum states} or simply {\it squeezed states}. In such quantum states, the uncertainty of the position and momentum operators are squeezed in some direction of the phase space, keeping their product satisfying the minimum uncertainty. Squeezed states occur in a wide variety of interesting scenarios, such as in quantum optics, the origin for the seed of large-scale structure formation in the form of the cosmological perturbations. The squeezing of a state is characterized by a squeezing parameter $r$ and a squeezing angle $\phi$. In the next section, we will describe the OTOC during the reheating phase from the squeezed initial state of the quantum fluctuations in the scalar fields governing the dynamics of the reheating phase.
Once we have the squeezing parameters, the momentum-space unequal time commutator is 
\begin{equation}
 \label{eq:mutc}
 [\hat{v}_k(\eta), \hat{\pi}_k(\eta_0)] = i(2\pi)^3\delta^3(\vec{k}-\vec{k}')\mathfrak{F}_k(\eta,\eta_0),
\end{equation}
with
\begin{align}
\nonumber
 \mathfrak{F}_k(\eta,\eta_0) =& \frac{1}{2}\Big[\left(\cosh r_k e^{-i\theta_k} - \sinh r_k e^{-i(\theta_k+2\phi_k)}\right)\\
 &\times\left(\cosh r_0 e^{-i\theta_k} + \sinh r_0 e^{-i(\theta_0+2\phi_0)}\right)+ c.c.\Big],
 \label{eq:unamp}
\end{align}
where the subscript $0$ is denoted to define a quantity evaluated at $\eta=\eta_0$, for instance: $r_0\equiv r_k(\eta_0)$. Finally, a Fourier mode OTOC as the double commutator is
\begin{equation}
    \mathcal{C}^{T}_{~\vk}(\eta) \equiv -\left\langle\left[\hat{v}_{\vk_1}(\eta),\hat{\pi}_{\vk_1'}(\eta_0)\right]\left[\hat{v}_{\vk_2}(\eta),\hat{\pi}_{\vk_2'}(\eta_0)\right]\right\rangle_{\beta},
    \label{eq:fk_sqz}
\end{equation}
where the two copies of the unequal time commutator with different momentum are used to avoid delta function divergences. Now, since the unequal time commutator is a $c$-number, after averaging, the OTOC reduces the square of the commutator
\begin{align}
\nonumber
    \mathcal{C}^{T}_{~\vk}(\eta) &\equiv -\left\langle\left[\hat{v}_{\vk_1}(\eta),\hat{\pi}_{\vk_1'}(\eta_0)\right]\left[\hat{v}_{\vk_2}(\eta),\hat{\pi}_{\vk_2'}(\eta_0)\right]\right\rangle\\
    &= (2\pi)^6 \mathfrak{F}_{k_1}(\eta, \eta_0)\mathfrak{F}_{k_2}(\eta, \eta_0)\delta^3(\vk_1 - \vk'_1)\delta^3(\vk_2 - \vk'_2).
    \label{eq:otoc_mom}
\end{align}
Barring the factors of $2\pi$ and the delta functions enforcing momentum conservation, the OTOC in~(\ref{eq:otoc_mom}) is proportional to the square of the amplitude of the unequal-time commutator defined in (\ref{eq:unamp}). Thus the OTOC and the amplitude $\mathfrak{F}_k$ have the same behavior for $\vk_1 \sim \vk_2 \sim \vk$. Thus following~\cite{Haque:2020pmp}, we will use the square of this amplitude as a prototype of the OTOC in our further analysis: 
\begin{equation}
    \mathcal{C}^{T}_{~\vk}(\eta) \equiv \abs{\mathfrak{F}_k(\eta, \eta_0)}^2.
    \label{eq:otoc_amp}
\end{equation}

\subsection{Quantum discord}
The concept of quantum discord was introduced to measure the \textit{quantumness of correlations} of two subsystems of a quantum system~\cite{Ollivier:2001fdq, Henderson2001:cqt}. 
A relatively close measure to quantum discord is the entanglement between the subsystems. However, quantum discord can be nonzero even if there is no entanglement, while zero discord implies entanglement is also zero. 
Consequently, quantum discord seems to be a better tool than quantum entanglement to look for nonclassical correlations in a system.
Moreover, when a system in the pure state is divided into two subsystems, the quantum discord is identical to the von Neumann entanglement entropy~\cite{bera2017quantum,datta2008quantum, Raveendran:2022dtb}. In defining quantum discord, we will use the original measurement-based version prevalent in the context of early Universe cosmology~(\cite{Lim:2014uea,Martin:2015qta,Hollowood:2017bil,Bhargava:2020fhl,Adhikari:2021pvv,Martin:2021znx,Raveendran:2022dtb}). (For other definitions of quantum discord, see, for instance,~\cite{bera2017quantum}. For formal development of complexity in quantum field theory, see, for instance, Refs.~\cite{Jefferson:2017sdb,Chapman:2017rqy,Caputa:2018kdj}).
One important detail about quantum discord worth mentioning here is that the quantum discord between two systems depends on which of the two systems is chosen for measurement, and hence, it is not necessarily symmetric. But when quantum discord coincides with entanglement entropy, i.e., when the total system is in a pure state, it is symmetric. The two-mode squeezed vacuum state is a simple example of entangled multimode field states providing rich phenomenology. The two-mode squeezing operator in (\ref{eq:sqz_op}) when acts on the two-mode vacuum state (initial state) $\ket{0}_{\vk}\ket{0}_{-\vk}$, yields
\begin{align}
\ket{\Psi_{\mathrm{sqz}}}_{\vk, -\vk} &= \mathcal{\hat{S}_{\vk}}(r_k, \phi_k) \ket{0}_{\vk}\ket{0}_{-\vk},\\
   						       &= \frac{1}{\cosh r_k}\sum_{n=0}^{\infty}(-1)^n e^{in\theta}(\tanh r_k)^n \ket{n_k,n_{-k}}
                \label{eq:sqz_state}
\end{align}
This state is an eigenstate of different number operators: $\hat{n}_k=\hat{c}^{\dagger}_{\vk}\hat{c}_{-\vk}$ and $\hat{n}_{-k}=\hat{c}^{\dagger}_{-\vk}\hat{c}_{\vk}$ with eigenvalue $0$. The average number of particles in each mode is the same as a result of strong correlation, i.e.,
\begin{align}
\langle\hat{n}_k\rangle = \langle\hat{n}_{-k}\rangle = \sinh^2 r_k,
\end{align}
The reduced density operators for the individual models are:
\begin{align}
\hat{\rho}_{k} &= \sum_{n=0}^{\infty} \frac{1}{(\cosh r_k)^2} (\tanh r_k)^{2n} \bra{n_k}\ket{n_k},\\
\hat{\rho}_{-k} &= \sum_{n=0}^{\infty} \frac{1}{(\cosh r_{-k})^2} (\tanh r_{-k})^{2n} \bra{n_{-k}}\ket{n_{-k}}
\label{eq:redensity}
\end{align}
while the probability of having n particles in a mode $k$ or $-k$ is
\begin{align}
p_n^{(i)} = \frac{(\tanh r_k)^{2n}}{(\cosh r_{k})^2},~\qquad~i=k,-k.
\end{align}
Notice that our two-mode squeezed state is already in the form of a Schmidt decomposition ($\ket{0}\ket{0}$); hence, the von Neumann entropy for our two-mode system is
\begin{align}
\nonumber
S(\hrho_k) &= -\trc[\hrho_k\ln\hrho_k] = S(\hrho_{-k} )\\
\nonumber
		&= -\sum_{n=0}^{\infty} p_n \ln p_n\\
		\nonumber
		&= \frac{(\tanh r_k)^{2n}}{(\cosh r_{-k})^2}\ln \frac{(\tanh r_k)^{2n}}{(\cosh r_{-k})^2}\\
		\nonumber
		&= \frac{(\tanh r_k)^{2n}}{(\cosh r_{-k})^2}\left[ \ln\left(\tanh^{2n}r_k\right) - \ln\left(\cosh^{2}r_k\right)\right]\\
		&= \cosh^2 r_k\ln(\cosh^2 r_k) - \sinh^2 r_k\ln(\sinh^2 r_k).
  \label{eq:von}
\end{align}

Finally, we will introduce a third measure of complexity, viz., the circuit complexity, in the section that follows.
\subsection{Circuit complexity}
As we saw in Sec.~\ref{sec:sqz_prim}, the curvature perturbations are naturally described by a system of two-mode squeezed states. 
We can define the complexity of this quantum circuit with the reference state being the two-mode vacuum state $\ket{0}_{\vk}\ket{0}_{-\vk}$ with the squeezed two-mode vacuum state $\ket{\Psi_{\mathrm{sqz}}}_{\vk, -\vk}$ serving as the target set. Following~\cite{Jefferson:2017sdb}, we define our states as Gaussian wave functions. Unlike Eqs. (\ref{eq:lagger1}) and (\ref{eq:lagger2}), here we define a set of auxiliary ``position'' and ``momentum'' variables with respect to the raising operator of $\vk$ instead as $-\vk$ as
\begin{equation}
    \hat{q}_{\vk}\equiv\frac{1}{\sqrt{2k}}\left(\hat{a}_{\vk} + \hat{a}^{\dagger}_{\vk}\right),\quad\hat{p}_{\vk}\equiv -i\sqrt{\frac{k}{2}}\left(\hat{a}_{\vk}-\hat{a}^{\dagger}_{\vk}\right)
\end{equation}
satisfying $\left[\hat{q}_{\vk},\hat{p}_{\vk}\right]=i\delta^3(\vk-\vk')$.
The Gaussian wave function describing the two-mode vacuum $\hat{a}_{\vk}\ket{0}_{\vk}\ket{0}_{-\vk}=0$ is:
\begin{align}
\nonumber
    \psi_R(q_{\vk},q_{-\vk}) &= \langle q_{\vk}|\langle q_{-\vk}|0\rangle_{\vk}|0\rangle_{-\vk}\\
                             &= \left(\frac{k}{\pi}\right)^{1/4}\exp(-\frac{k}{2}\left(q^2_{\vk}+q^2_{-\vk}\right)).
\end{align}
To find the wave function corresponding to the squeezed state in (\ref{eq:sqz_state}), we note that the following combination annihilates it:
\begin{equation}
    \left(\cosh r_k \hat{a}_{\vk} + e^{-2i\phi_k}\sinh r_k \hat{a}^{\dagger}_{\vk}\right)\ket{\Psi_{\mathrm{sqz}}}_{\vk, -\vk} = 0.
\end{equation}
This fetches us with the following form of the ``position-space'' wave function~\cite{Martin:2019wta}:
\begin{align}
  \ket{\Psi_{\mathrm{sqz}}}_{\vk, -\vk} = \frac{\exp(A(q^2_{\vk}+q^2_{-\vk})-Bq_{\vk}q_{-\vk})}{\sqrt{\pi}\cosh r_k\sqrt{1-\tanh^2r_k\exp(-4i\phi_k)}}  
  \label{eq:psi_sqz}
\end{align}
where the coefficients $A$ and $B$ are defined as
\begin{align}
\label{eq:coffA}
    A &= \frac{k}{2}\left(\frac{\tanh^2r_k\exp(-4i\phi_k)+1}{\tanh^2r_k\exp(-4i\phi_k)-1}\right),\\
    B &= 2k\left(\frac{\tanh^2r_k\exp(-2i\phi_k)}{\tanh^2r_k\exp(-4i\phi_k)-1}\right),
    \label{eq:coffB}
\end{align}
Before we write the working definition of circuit complexity, a few words on the different methods of computing the quantity between a reference and a target state are in order. Among different methods, Nielsen's geometric method for computing the complexity defines it to be \textit{equivalent to finding the shortest path between two points in a certain curved geometry}~\cite{Nielsen:2006qcg}. Additionally, rather than using the wave function for computing circuit complexity as in (\ref{eq:psi_sqz}), we can also use the covariance matrix. It has been shown in~\cite{Bhattacharyya:2020rpy} that the two definitions will result in different forms of circuit complexity. Notably, the circuit complexity from the covariance metric will be independent of the squeezing angle. This will, in turn, imply that the circuit complexity will grow even during non--de Sitter expansion (cf. the discussion in the next section). However, as pointed out in~\cite{Ali:2019zcj}, the circuit complexity computed using the covariance matrix is less sensitive to the fine details of the states, and consequently, the definition using the wave function of the target state is a more precise measure of the circuit complexity.
\par
Now using Nielsen's geometric and employing the \textit{geodesic weighting}~\cite{Jefferson:2017sdb}, we obtain the expression for circuit complexity $\mathcal{C}$ as:
\begin{widetext}
\begin{equation}
    \mathcal{C}(k) = \frac{1}{2}\sqrt{\left(\ln\left|\frac{\Omega_{\vk}}{\omega_{\vk}}\right|\right)^2 + \left(\ln\left|\frac{\Omega_{-\vk}}{\omega_{-\vk}}\right|\right)^2 + \left(\tan^{-1}\frac{\operatorname{\mathbb{R}e}\Omega_{\vk}}{\operatorname{\mathbb{I}m}\Omega_{\vk}}\right)^2 + \left(\tan^{-1}\frac{\operatorname{\mathbb{R}e}\Omega_{-\vk}}{\operatorname{\mathbb{I}m}\Omega_{-\vk}}\right)^2}
    \label{eq:nielsenC}
\end{equation}
\end{widetext}
where $\Omega_{\pm\vk}=-2A\pm B$, and $\omega_{\vk}=\omega_{-\vk}=k/2$. Now using (\ref{eq:coffA}) and (\ref{eq:coffB}) in (\ref{eq:nielsenC}), we obtain the following closed-form expression for complexity for the two-mode squeezed vacuum state relative to the reference:
\begin{widetext}
\begin{equation}
    C(k) = \frac{1}{\sqrt{2}}\sqrt{\left|\ln\left|\frac{1+\tanh r_k\exp(-2i\phi_k)}{1-\tanh r_k\exp(-2i\phi_k)}\right|\right|^2 + \abs{\tan^{-1}(2\sin2\phi_k\sinh r_k\cosh r_k)}^2}.
    \label{eq:comp1}
\end{equation}
\end{widetext}
\par
By now, we have the basic definitions of the three complexity measures of our interest. However, it is hard to decipher their significance for our system in these forms. To guide our intuition before fully analyzing the system numerically, we will now try to understand the behavior of the squeezed parameters and these complexity measures by taking some ansatz for the evolution of the scale factor under different expansion histories of our Universe. We present this analysis in the next section.
\section{\label{sec:expanding_univ}Complexity in the Expanding Background}
The evolution of the squeezing parameters and different complexity measures using some ansatz for the scale factor evolution are well studied in the literature. The behavior for both expanding and contraction backgrounds can be found in~\cite{Bhattacharyya:2020rpy,Bhattacharyya:2020kgu}. For our present purpose, i.e., the evolution of the primordial complexity in our Universe, we will only consider the expanding backgrounds. By and large, the expansion history during different epochs in our Universe is dominated by a single component. Hence, the energy dilution due to the expansion of the Universe dominated by ideal fluid satisfying $p = w\rho$ is aptly described by $\rho \propto a^{-3(1+w)}$. Now defining $\beta=-2/(1+3w)$, we write the following ansatz for the scale factor evolution:
\begin{equation}
    a(\eta) = \left(\frac{\eta_0}{\eta}\right)^{\beta} = 
    \begin{cases}
         &~\text{accelerating background},\\
         \left(\frac{\eta_0}{\eta}\right)^{\beta}& -\infty<\eta<0,\eta_0<0\\ 
                        & \beta>0~(w<-1/3),\\
                        &\\
          &~\text{decelerating background},\\
         \left(\frac{\eta_0}{\eta}\right)^{|\beta|}& 0<\eta<\infty,\eta_0>0\\ 
                        & \beta>0~(w>-1/3),\\
         \end{cases}
\end{equation}
To analyze the behavior of squeezing parameters, it is useful to substitute the scale factor as the independent variable. The differential equations governing the evolution of squeezing parameters given in~(\ref{eq:rk}-\ref{eq:thetak}) now reduce:
\begin{align}
\label{eq:rk_approx0}
    \frac{\dx r_k}{\dx a} &= -\frac{1}{a}\cos(2\phi_k),\\
    \label{eq:phik_approx0}
    \frac{\dx \phi_k}{\dx a} &= -\frac{k|\eta_0|}{|\beta|}\frac{1}{a^{1+1/\beta}} + \frac{1}{a}\coth{2r_k}\sin{2\phi_k},\\
       \label{eq:thetak_approx0}
    \frac{\dx \theta_k}{\dx a} &= \frac{k|\eta_0|}{|\beta|}\frac{1}{a^{1+1/\beta}} - \frac{1}{a}\tanh{2r_k}\sin{2\phi_k}.
\end{align}
The solutions to Eqs. (\ref{eq:rk_approx0}-\ref{eq:thetak_approx0}) depend on whether the background expansion is accelerating or decelerating. We will take care of the two scenarios separately. 

\subsection{Accelerating background}
For accelerating backgrounds ($w<-1/3$) that satisfy the \textit{null energy condition}, the background EOS follows $-1\leq w \leq -1/3$. The derived parameter $\beta$ in such cases has the range $1 \leq \beta < \infty$. One such spacetime example is the inflationary Universe starting from the de Sitter background $w=-1$ ($\beta=1$) to the end of inflation when $w=-1/3$. The exact solutions to the squeezing parameters are known for the de Sitter background given in~(\ref{eq:bdsqz1})-(\ref{eq:bdsqz3}). For accelerating expansion, at sufficiently early times $a \ll 1$, all perturbation modes start inside the horizon. If the modes start with small squeezing $r_k \ll 1$ and constant squeezing angle, the solutions to Eqs. (\ref{eq:rk_approx0})-(\ref{eq:thetak_approx0}) is
\begin{align}
\label{eq:sol_rk1}
    r_k(a) &\approx \frac{\beta}{2k|\eta_0|}a^{1/\beta} \ll 1,\\
 \label{eq:sol_phik1}
    \phi_k(a) &\approx \frac{\pi}{4} - \frac{\beta}{2k|\eta_0|}a^{1/\beta},\\
    \label{eq:sol_thetak1}
    \theta_k(a) &\approx - \frac{2k|\eta_0|}{a^{1/\beta}} \gg 1.
\end{align}
Now taking the initial state to be unsqueezed $r_0 \to 0$, the unequal-time commutator amplitude in (\ref{eq:unamp}) reduces to:
\begin{align}
\nonumber
    \mathfrak{F}_k(\eta,\eta_0) =& \cosh r_k \cos(\theta_k - \theta_0)\\
    &- \sinh r_k \cos(\theta_k - \theta_0 + 2\phi_k).
    \label{eq:mod_amp_fk}
\end{align}
It is also clear from Eq.(\ref{eq:comp1}) that for the accelerating FLRW backgrounds, at early times, the complexity vanishes $\mathcal{C}\to 0$ for all equation of state parameters.
\newline
~~At times well after the initial time of the evolution but until the modes are within the horizon, the amplitude function given in (\ref{eq:mod_amp_fk}) oscillates with unit amplitude $\mathfrak{F}_k(\eta,\eta_0) \approx \cos(\theta_k - \theta_0)$. Therefore, at early times, for OTOC defined in (\ref{eq:otoc_amp}) we have $\mathcal{C}^{T}_{~\vk}\sim\mathcal{O}(1)$. On the other hand, the leading order solution at late times when the modes exit the horizon ($a \gg (k|\eta_0|)^{\beta}$) is given by~\cite{Bhattacharyya:2020kgu}
\begin{align}
\label{eq:sol_rk2}
    r_k(a) &\approx \ln(\frac{a}{k|\eta_0|})\\
 \label{eq:sol_phik2}
    \phi_k(a) &\approx \frac{\pi}{4},\\
    \label{eq:sol_thetak2}
    \theta_k(a) &\approx - \frac{2k|\eta_0|}{2\beta-1}\frac{1}{a^{1/\beta}}\ll 1.
\end{align}
i.e., once the modes exit the horizon, the squeezing parameter, $r_k$ grows while the rotation angles freeze out to constant values. Consequently, the late-time unequal-time commutator is determined by the large squeezing parameter: $\mathfrak{F}_k(\eta,\eta_0) \approx \frac{a(\eta)}{k|\eta_0|}\cos\theta_0$. Finally, the late-time behavior of OTOC is $\mathcal{C}^{T}_{~\vk}\sim\left(\frac{a(\eta)}{k|\eta_0|}\right)^2\cos\theta_0$. On the other hand, at late times, the first term in (\ref{eq:comp1}) dominates over the second, yielding
\begin{equation}
    \mathcal{C} = \frac{1}{\sqrt{2}}\Bigg|\ln{\Big|\frac{1+e^{-2i\phi_k\tanh{r_k}}}{1-e^{-2i\phi_k\tanh{r_k}}}}\Big|\Bigg|
    \label{eq:approx_com}
\end{equation}
As at this stage, $r_k\gg 1,~\tanh{r_k}\approx1$, the dependence of complexity in (\ref{eq:approx_com}) on the scale factor is mainly due to the squeezing angle $\phi_k$. Using the approximate solutions for the squeezing parameters, we can see that the complexity in the accelerating background will increase indefinitely until the background changes and the expansion becomes nonaccelerating.
The behavior of the von Neumann entropy readily follows the behavior of the squeezing parameter as $S(\hrho_k)\approx r_k$.

\begin{figure*}
\centering
\includegraphics[width=0.75\textwidth]{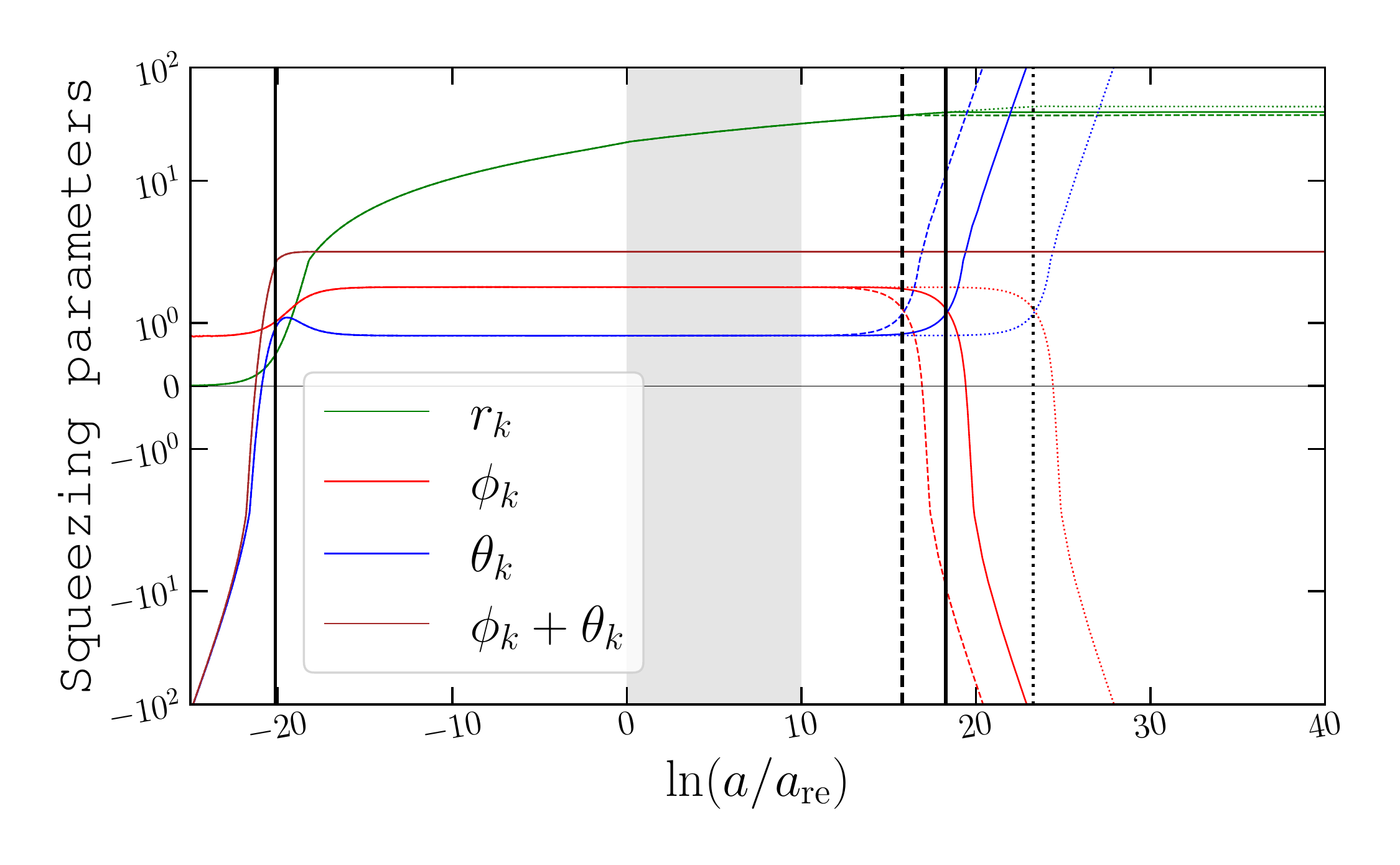}
\caption{We plot the evolution of the squeezing parameter $r_k$, $\phi_k$, and $\theta_k$. The shaded region is the reheating phase (taken to be $\nre=10$). The solid lines are for the instant reheating ($\wre=1/3$), while the dotted and dashed curves are for the case of $\wre=<1/3$ and $\wre>1/3$, respectively. The fiducial mode considered here exits the horizon at around $20$ $e$-folds before the end of inflation, denoted by the solid vertical line. The horizon reentries for the mode (assuming a fixed reheating epoch ) are denoted as the vertical dotted line (for $\wre=0$), vertical solid line (for $\wre=1/3$), and vertical dashed line (for $\wre>1/3$).}
\label{fig:fisqzcurv}
\end{figure*}
\begin{figure*}[ht!] 
  \includegraphics[width=0.42\textwidth]{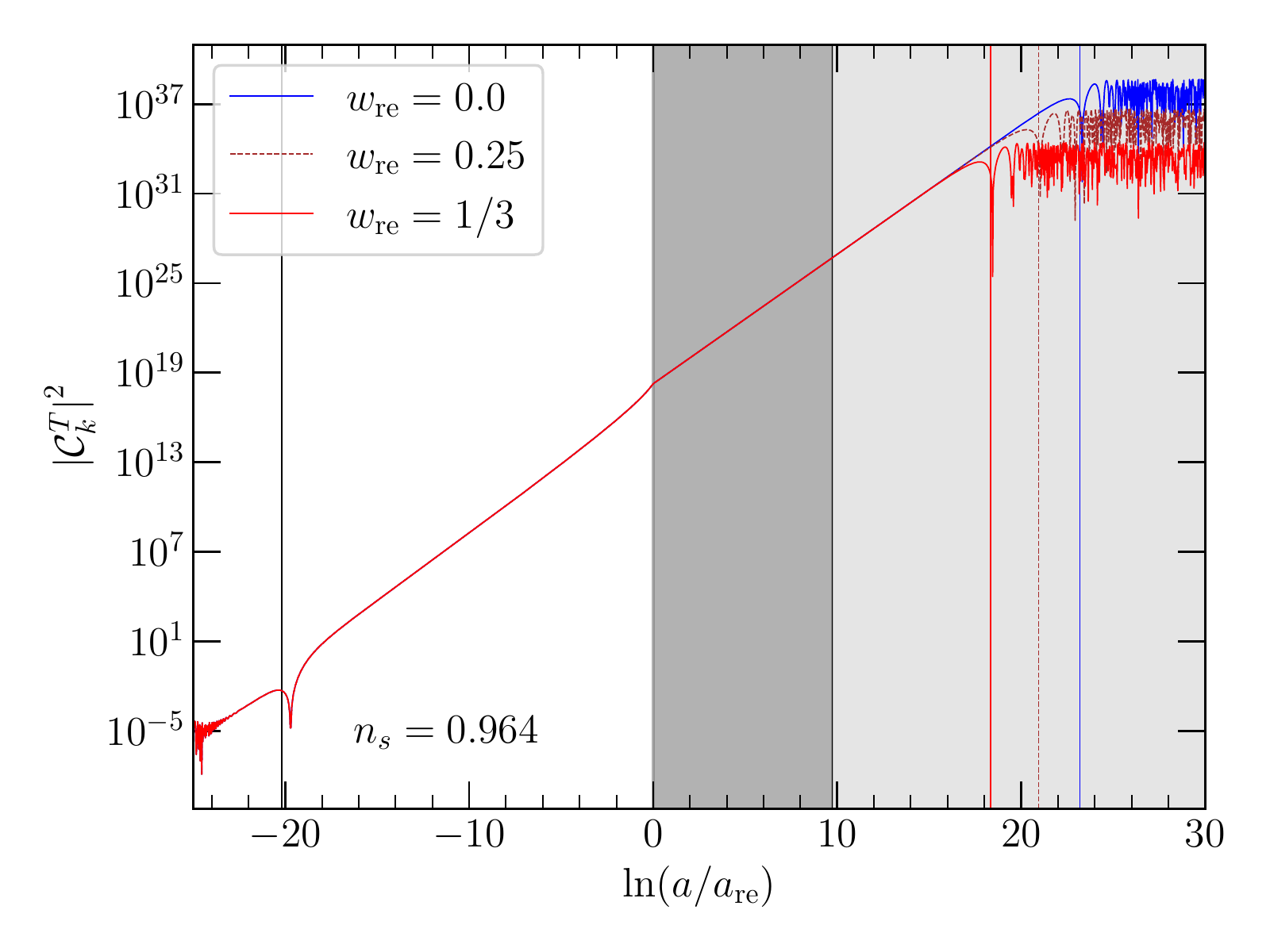}~\includegraphics[width=0.42\textwidth]{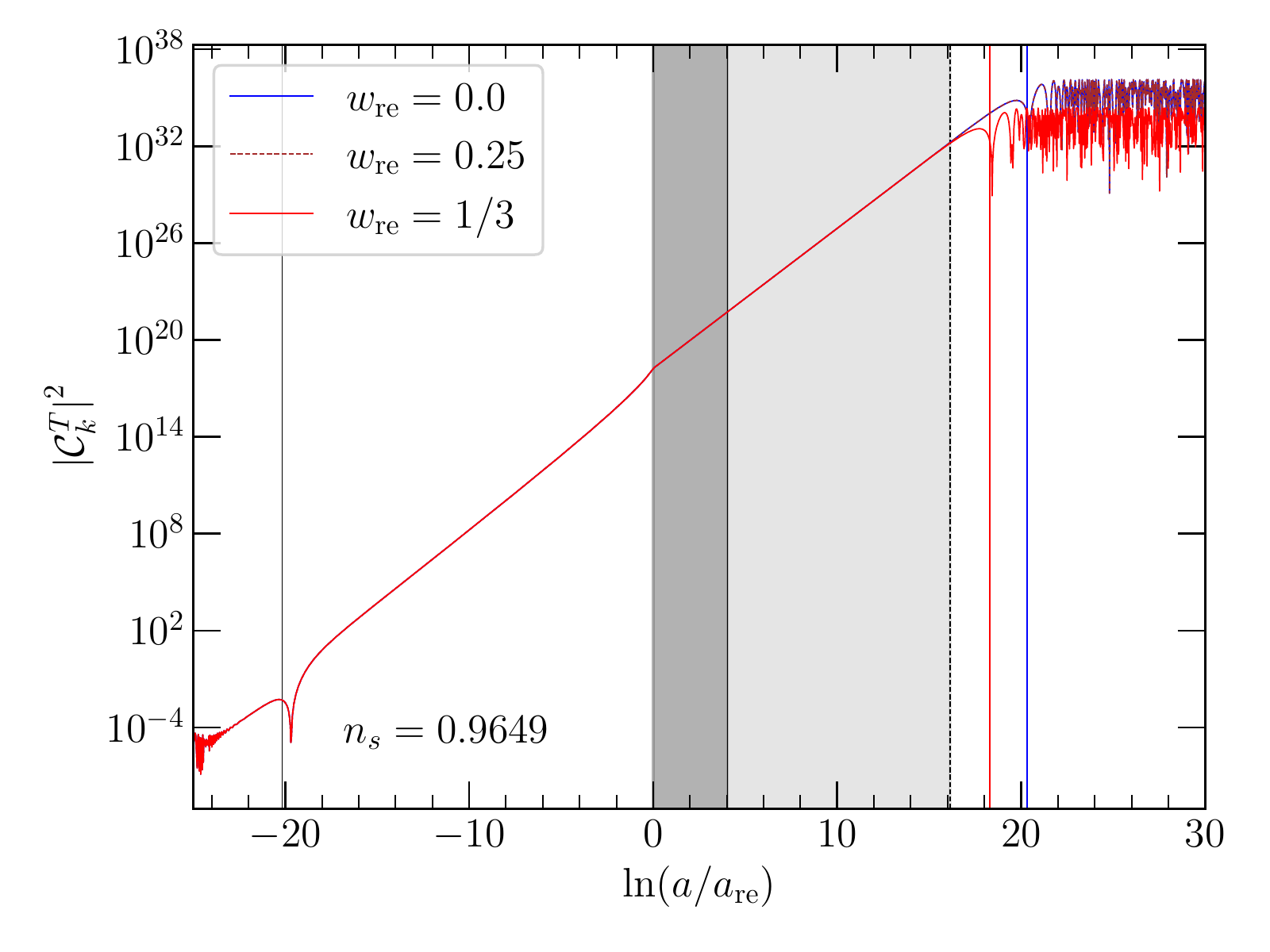}\\%
  \includegraphics[width=0.42\textwidth]{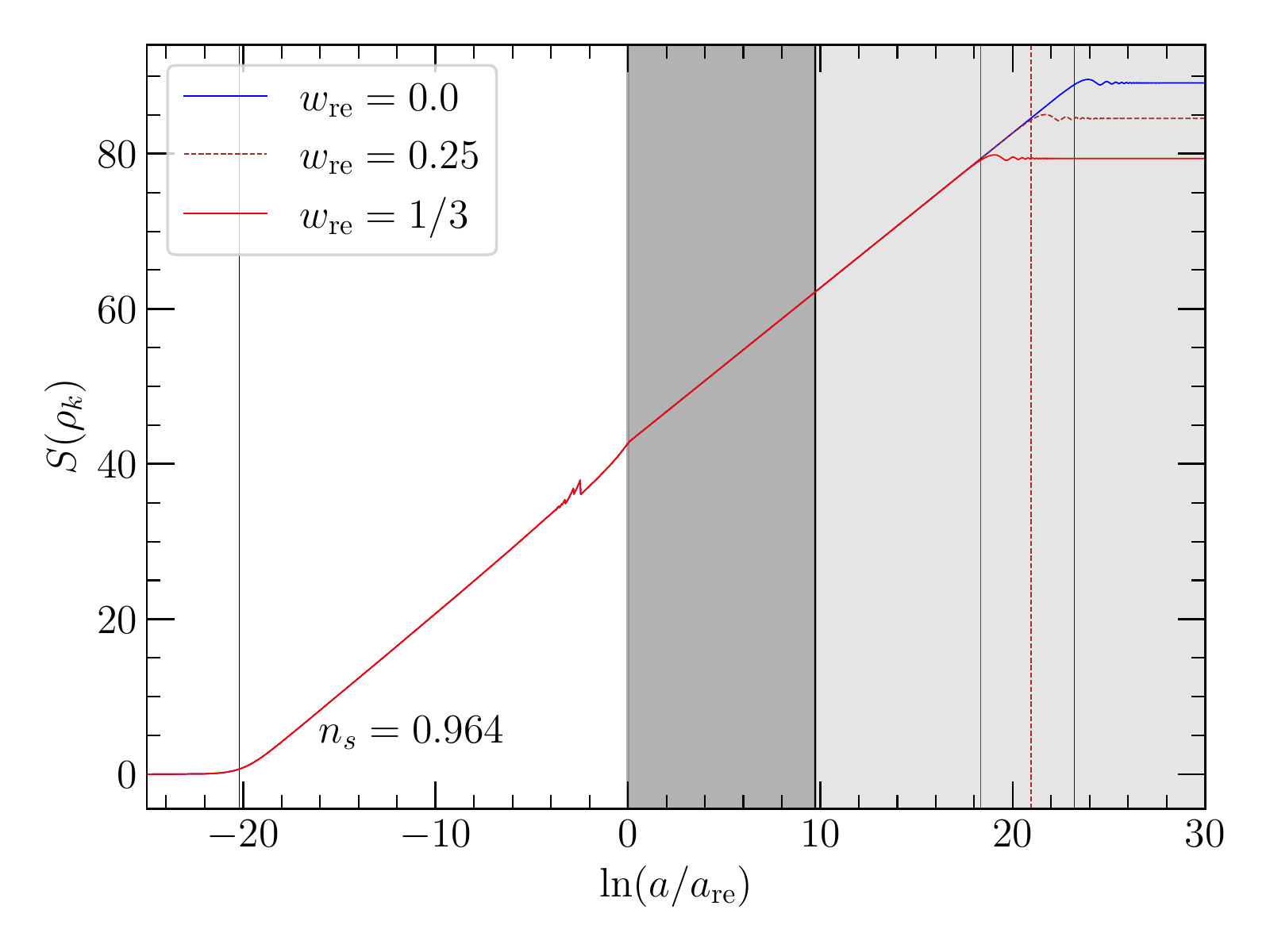}~\includegraphics[width=0.42\textwidth]{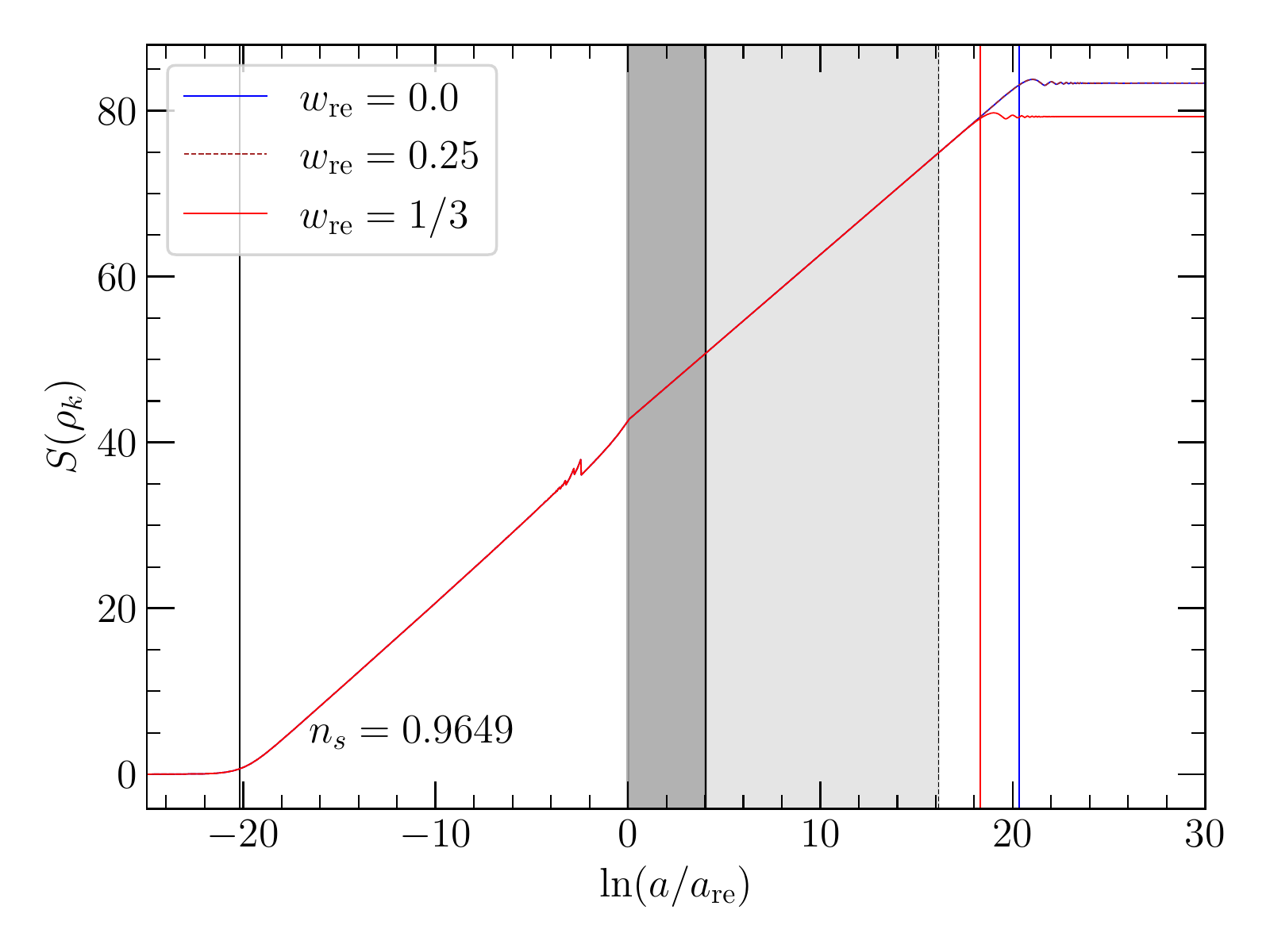}
\caption{Evolution of the OTOC and quantum discord for a reference mode for two different values of the EOS parameter. The evolution of the correlations follows the pattern of the squeezing parameters. The OTOC and the von Neumann entropy grow when the modes are outside the horizon. Upon entering the horizon, they settle to a fixed amplitude. 
The red curves are when the Universe goes through instantaneous reheating. The top figures are for the central value of $n_s=0.9649$ from Planck data. 
The duration of the reheating epoch for the two different EOS parameters is shown as the shaded regions (darker region for $\wre=0$). 
For these specific reheating histories, the reference mode renters the horizon during the radiation-dominated era~---~consequently, the individual reheating EOS has no effect on the evolution of the correlations. 
On the other hand, for $n_s=0.964$, the reference mode will reenter the horizon during the reheating epoch when $\wre=0.25$. In this case, the signature of the EOS parameters will be on the evolution of the correlations. The solid black line denotes the horizon exit, while the reentry is denoted by the vertical red~($\wre=1/3$) and solid blue ~($\wre=0$) and blue dashed ($\wre=0.25$), respectively.}
\label{fig:evolve_ccomp1}
\end{figure*}
\begin{figure*}[h!] 
  \includegraphics[width=0.42\textwidth]{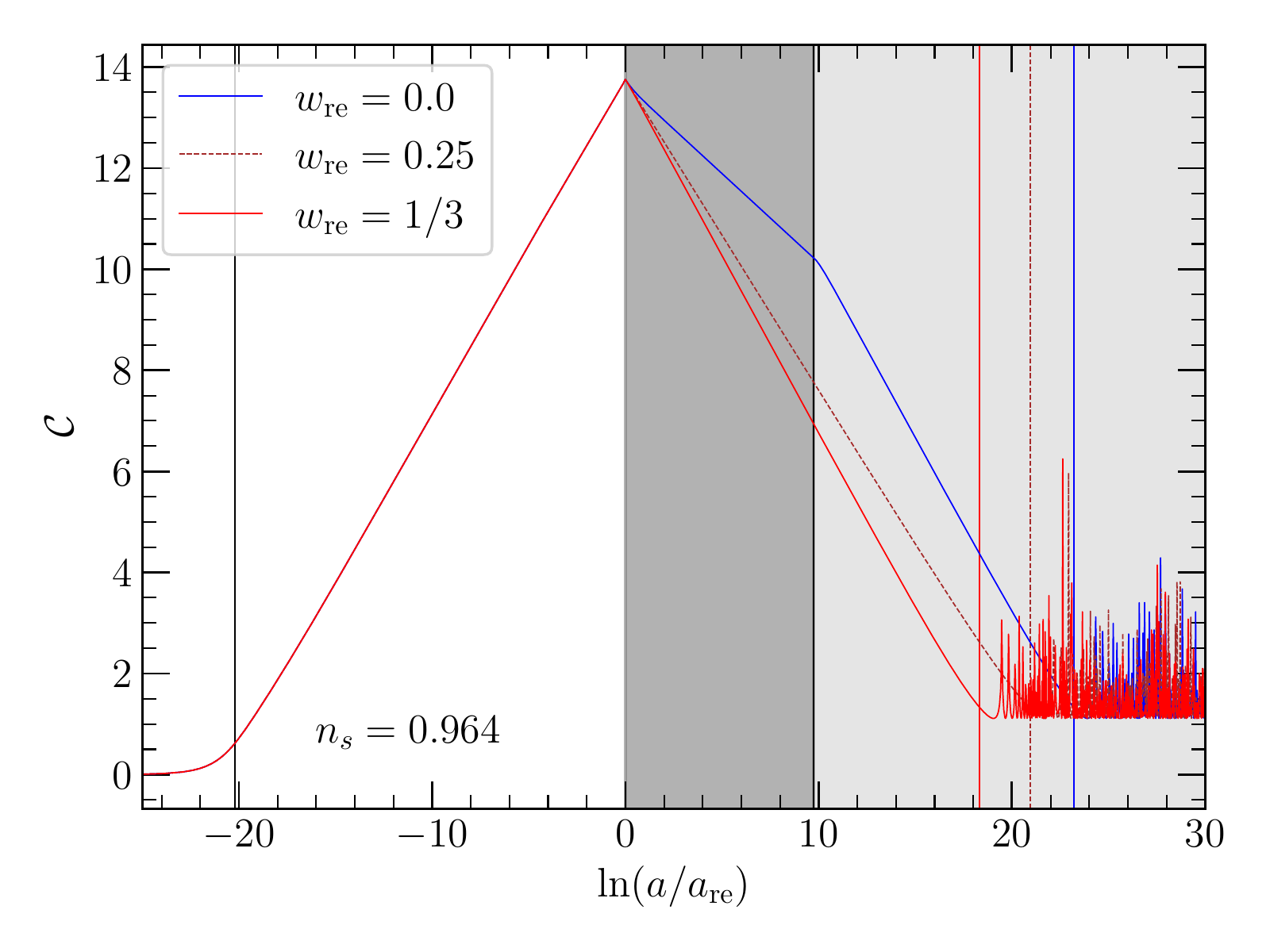}~\includegraphics[width=0.42\textwidth]{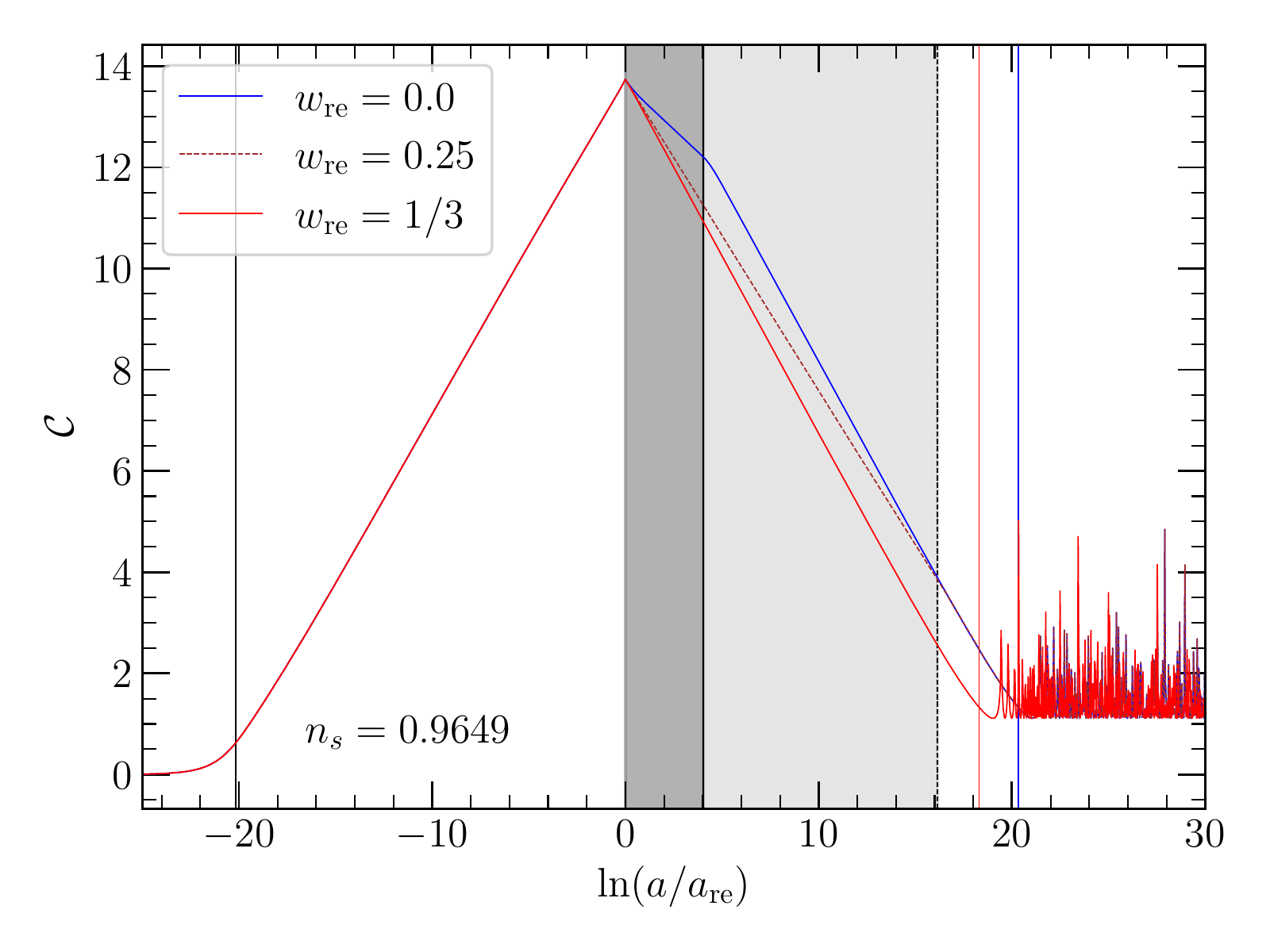}\\
\includegraphics[width=0.42\textwidth]{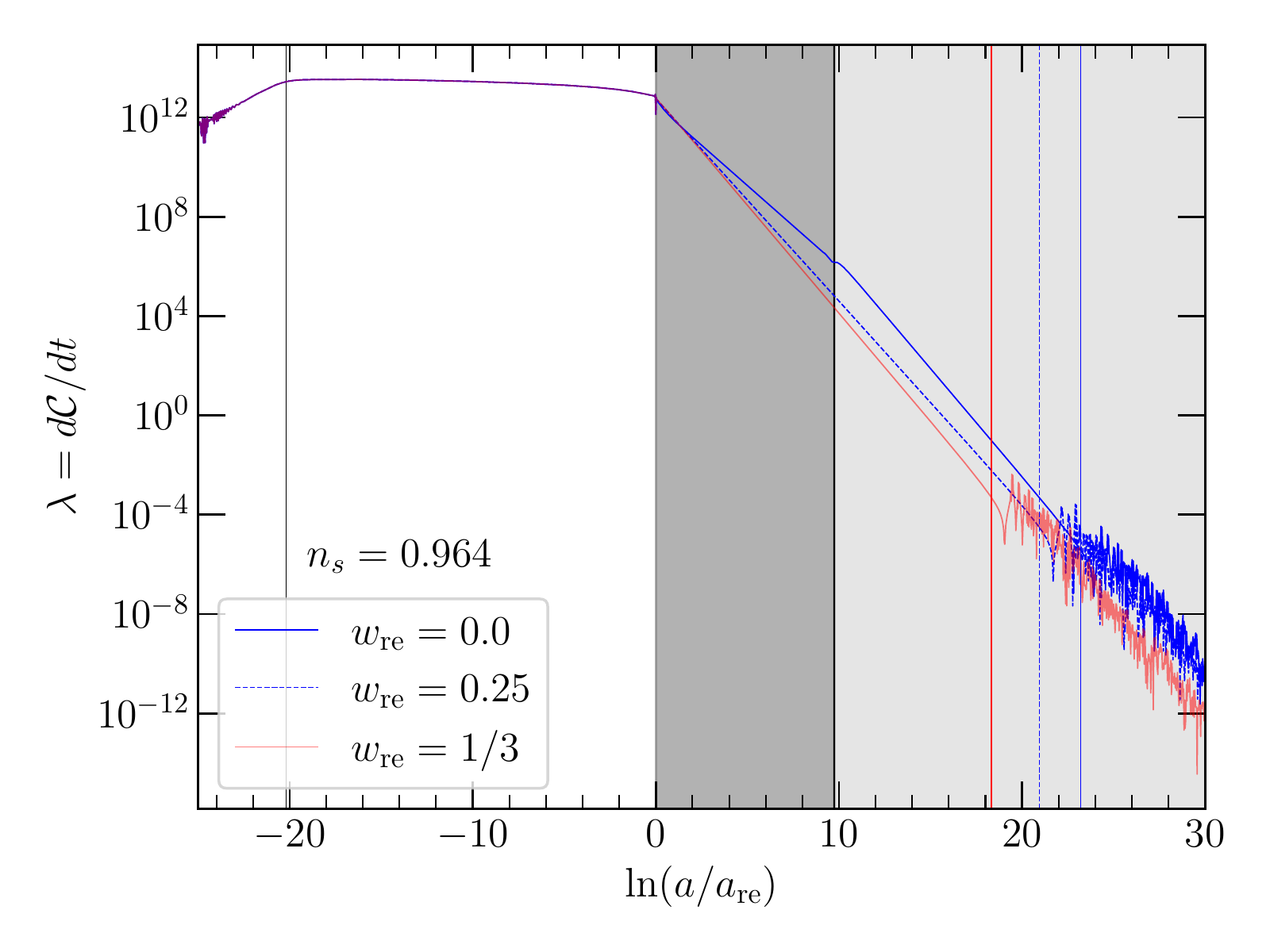}~\includegraphics[width=0.42\textwidth]{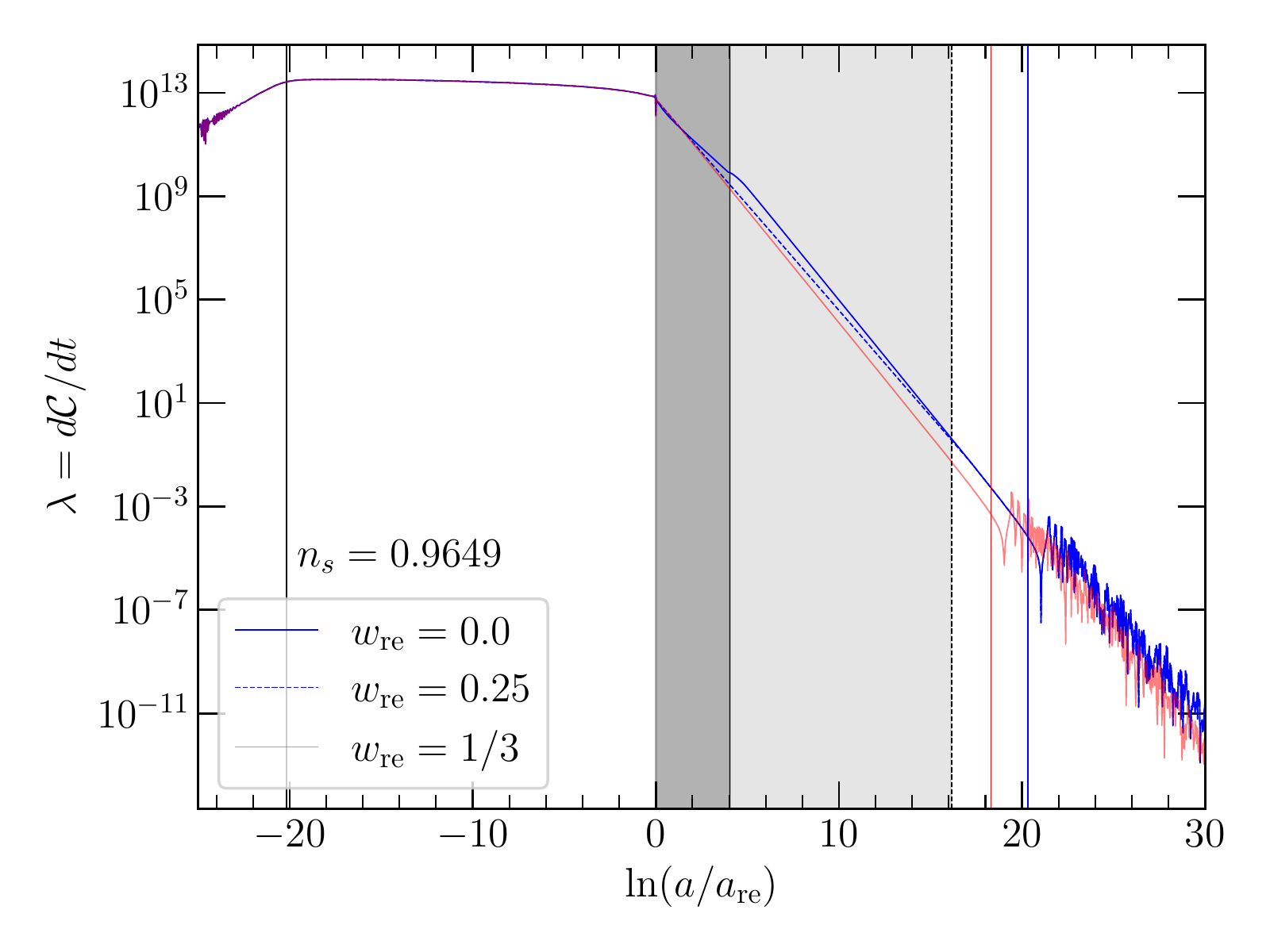}
\caption{Evolution of the circuit complexity (top panel) and the slope of the circuit complexity, which is a measure of the Lyapunov exponent $\lambda$. Contrary to the previous two complexity measures (as summarized in Fig. \ref{fig:evolve_ccomp1}), the different reheating epochs directly affect the evolution circuit complexity.}
\label{fig:evolve_ccomp2}
\end{figure*}
\subsection{Decelerating backgrounds}
We now consider the decelerating backgrounds $w>-1/3$. The expansion of the Universe after the end of inflation until the recent dark energy-dominated expansion is described by the decelerating expansion. In this case, the modes begin outside the horizon for sufficiently early times $a\ll 1/(k\eta_0)^{|\beta|}$. Naturally, the solutions in this case qualitatively resemble the late-time superhorizon solutions in the previous accelerating case.
\begin{align}
\label{eq:sol_rk3}
    r_k(a) &\approx r_0 + \ln(\frac{a}{|a_0|})\\
 \label{eq:sol_phik3}
    \phi_k(a) &\approx \frac{\pi}{2} - \frac{k\eta_0}{|\beta|(|\beta|+2)}a^{1/\beta},\\
    \label{eq:sol_thetak3}
    \theta_k(a) &\approx \frac{k\eta_0}{|\beta|(|\beta|+2)}a^{1/\beta}\ll 1.
\end{align}
Now taking the large initial squeezing $r_0\gg 1$, and their asymptotic vales of the squeezing angles $\phi_0\approx\pi/2$ and $\theta_0\approx 0$, the amplitude function is:
\begin{equation}
    \mathfrak{F}_k(\eta,\eta_0) = e^{-r_0}\left(\cosh r_k \cos \theta_k - \sin r_k \cos(\theta_k + 2\phi_k)\right)
    \label{eq:amp_fk1}
\end{equation}
Using the above solutions for squeezing parameters, the amplitude function in (\ref{eq:amp_fk1}) reduces to
\begin{equation}
    \mathfrak{F}_k(\eta, \eta_0) \approx e^{r_k - r_0} \approx \frac{a(\eta)}{a_0} \geq 1.
\end{equation}
Thus, the OTOC for a mode increases with the scale factor during the early times until the horizon crossing. 
\newline
Finally, when a highly squeezed mode reenters the horizon at late times $a \gg 1/(k\eta_0)^{\abs{\beta}}$, the equation for $\phi_k$ becomes
\begin{equation}
\frac{\dx \phi_k}{\dx a} = -\frac{k|\eta_0|}{|\beta|}\frac{1}{a^{1+1/\beta}}
\end{equation}
The solution is a decaying function of the scale factor (similarly, the $\theta_k$ is a growing function of $a$). Correspondingly, the equation of $r_k$ will have an oscillatory term in the rhs that will put off its growth. Thus, for a highly squeezed mode, the squeezing parameter will \textit{`freeze in'} to the value of the squeezing at the horizon crossing upon horizon reentry~\cite{Bhattacharyya:2020rpy}.
\begin{align}
    r_k(a)    &\approx r_{\ast},\\
    \phi_k(a) &\approx -k\eta_0 a^{1/|\beta|},\\
    \theta_{k}(a) &\approx k\eta_0 a^{1/|\beta|}
\end{align}
Consequently, the OTOC will oscillate with an amplitude set by the scale factor at horizon reentry $a_{\ast} \sim 1/k\eta_0$,
\begin{align}
\nonumber
   \mathfrak{F}_k(\eta,\eta_0) &\approx e^{r_0}\left(\cosh r_{\ast}\cos\theta_k + \sinh r_{\ast}\cos(\theta_k + 2\phi_k)\right)\\
   &\approx \frac{a_{\ast}}{a_0}\cos(k\eta_0a^{1/|\beta|}),
\end{align}
The behavior of the complexity, in this case, is studied using the late-time expression given in (\ref{eq:approx_com}) as we have $r_k\gg 1$. However, $\phi_k$ is drifting away from a constant value at this stage. In this case, the circuit complexity is found to be dependent on the equation of the state parameter as
\begin{equation}
    \mathcal{C} \propto -\frac{1+3w}{2\sqrt{2}}\ln{a}.
\end{equation}
and consequently, the rate of change of $\mathcal{C}$ with respect to cosmic time is
\begin{equation}
   \frac{d\mathcal{C}}{dt} \approx -\frac{1+3w}{2\sqrt{2}}H.
\end{equation}
This is an interesting result as the other quantities that we defined, such as the OTOC or the quantum discord, do not bear any direct signature of the equation of state during the evolution phase when the Universe is going through a nonaccelerating expansion (the reheating phase after inflation and the RD or MD). Furthermore, a period of an approximately constant phase of complexity followed by a phase of linear growth~---~as we observed in our case~---~is shown to be a characteristic of quantum chaos in a system~\cite{Ali:2018fcz,Ali:2019zcj}. The study of chaos via quantum correlators often involves defining two parameters: the \textit{scrambling time} and the \textit{Lyapunov exponent}, and this motivates the authors in~\cite{Bhattacharyya:2020kgu} to conjecture that the duration of the approximately constant part of the complexity is equivalent to the scrambling time while the slope in the linear portion is a measure of the Lyapunov exponent ($\lambda=d\mathcal{C}/dt$). Consequently, these results can be used to understand the development of chaos in the early Universe. 
Now, with the analytic results to guide our intuition, we will check the results with the full numerical solution in the cosmic background in the early Universe---starting from the inflationary Universe and then reheating epoch, and finally in the radiation-dominated era.
\begin{figure*}
     \centering
    \includegraphics[width=0.42\textwidth]{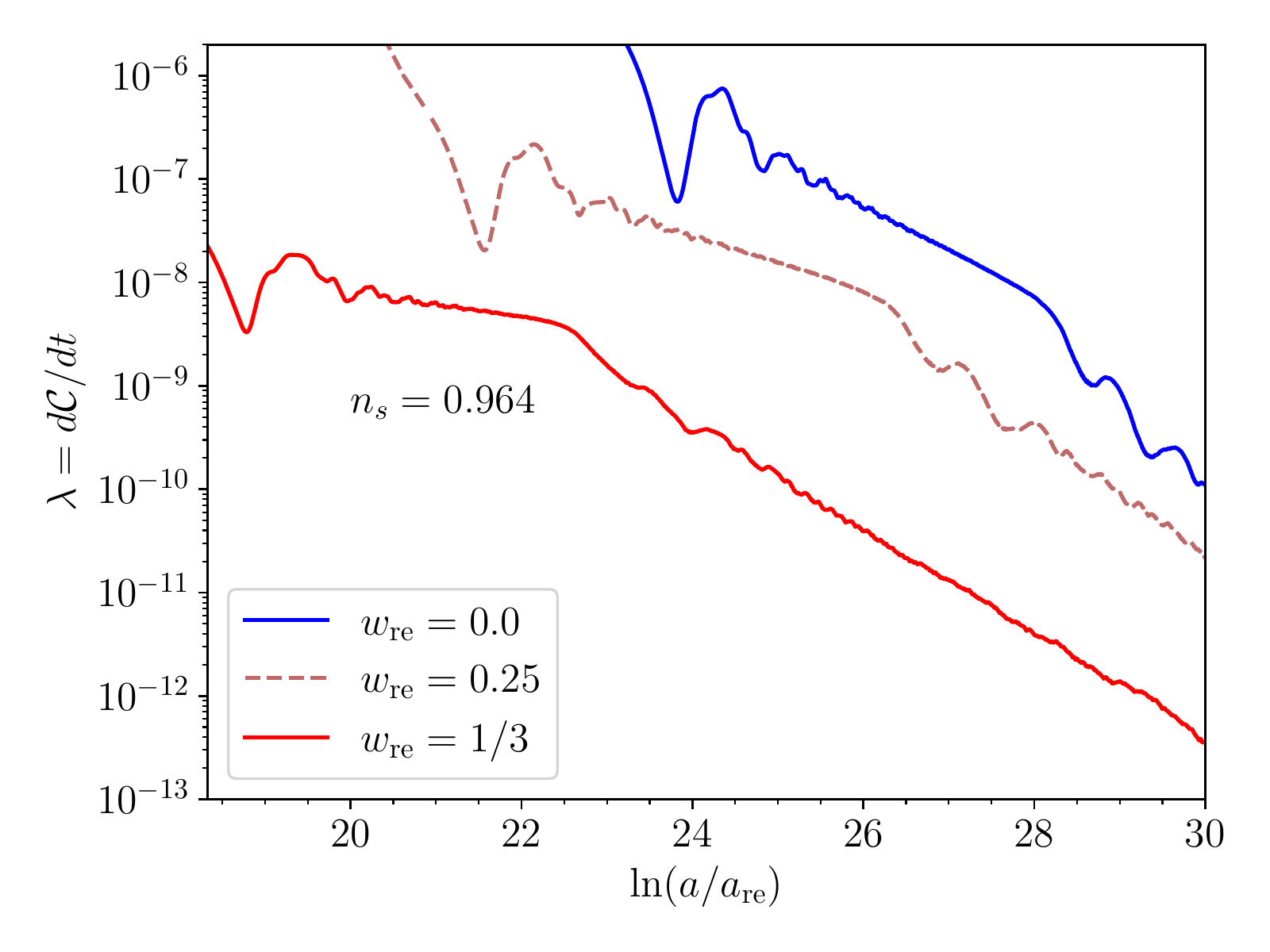}
    \includegraphics[width=0.42\textwidth]{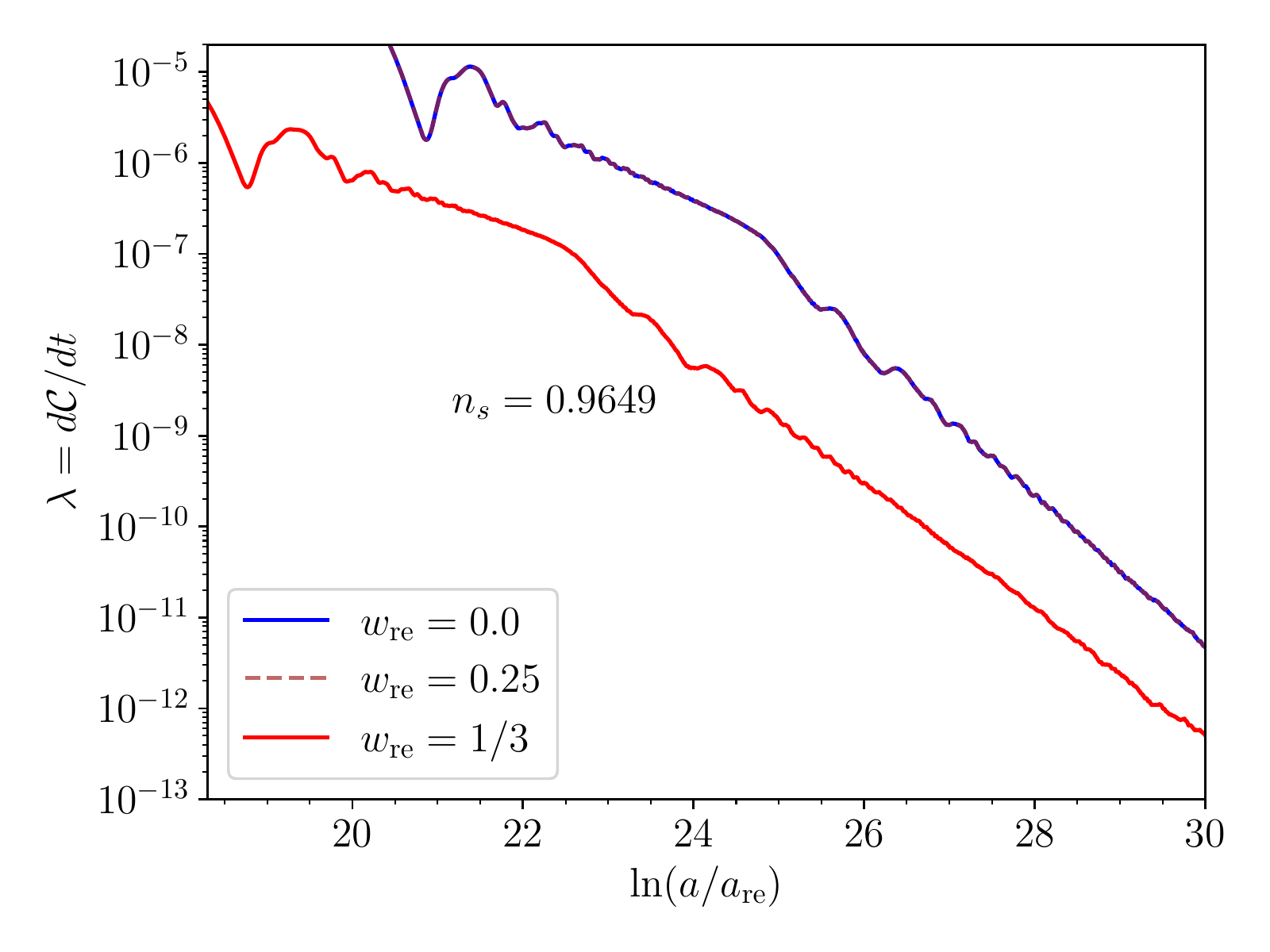}
    \caption{Evolution of the slope of circuit complexity $d\mathcal{C}/dt$ after horizon reentry. 
    Although they evolve differently during different reheating epoch, the final value follows the same evolution pattern as before.}
     \label{fig:evolve_ccomp3}
\end{figure*}
\section{\label{sec:res_dis}Results and Discussion}
The primordial complexity measures defined in (\ref{eq:unamp})-(\ref{eq:otoc_amp}), (\ref{eq:von}), and (\ref{eq:comp1}) are completely specified by the squeezing parameters.
Ergo, our natural starting point would be to understand the evolution of squeezing parameters. 
We have seen the approximate analytic behavior of the quantities in the previous section that the squeezing parameter $r_k$ grows when the perturbation modes are outside the horizon and eventually \textit{freezes in} to a constant value at the horizon reentry. 
The squeezing angles remain approximately constant inside the horizon and grow with opposite signs after the horizon exit.
\par
Now, to understand the evolution of primordial complexity without any approximations, we will numerically integrate the squeezing equations in (\ref{eq:rk}-\ref{eq:thetak}) with the initial conditions given in (\ref{eq:bdsqz1}-\ref{eq:bdsqz3}).
The results are plotted in Fig~(\ref{fig:fisqzcurv}) for three cases of reheating history, namely $\wre=0$, $1/3$, and $1/2$. 
In plotting the squeezing parameters, our prime objective is to understand the effects of the reheating epoch on the evolution of the squeezing parameters. Consequently, we took a fixed reheating epoch ($\nre=10$) in all the cases. 
As we saw in our previous discussions, the reheating phase only affects the evolution of the parameters in deciding when the modes reenter the horizon. 
In general, the squeezing parameter $r_k$ grows when they are outside the horizon. 
The phase parameters $\theta_k$ and $\phi_k$ grow inside the horizon while the combination $\phi_k + \theta_k$ remains constant.  
For studying the evolution of the squeezing parameters, we took the reheating history consistently. 
For this, we first fixed the scalar spectral index. 
As we saw from the discussion in Sec (\ref{sec:reh_cons}), once we know the spectral index, the evolution of the Hubble horizon is known modulo the uncertainty of the reheating epoch, which we have modeled as the expansion of the Universe dominated by an effective fluid with EOS parameter $\wre$. 
\par
Now, we are in a position to study the evolution of the complexity measures. 
For this, we will follow their evolution in two different evolutionary histories of the Universe determined by the spectral index $n_s$. 
Remember that once we fix $n_s$, given an inflationary model, we know the reheating $e$-folds $\nre$, and hence the background evolution of the Universe is completely specified.
We showed the evolution of OTOC and Quantum discord (the von Neumann entropy) in Fig \ref{fig:evolve_ccomp1}. 
The evolution of the circuit complexity and its derivative with respect to the cosmic time is shown in Fig \ref{fig:evolve_ccomp2}.
The upper panel figures are for $n_s=0.9649$ when the CMB pivot scale exits the horizon at $N_k=56.48$. 
In this case, the reheating $e$-folds when $\wre=0.0$ and $0.25$ are $\nre=4$ and $\nre=16$, respectively. 
The red curve in the plots denotes the evolution with instantaneous reheating. 
The blue solid ($\wre=0.0$) and brown dashed ($\wre=0.25$) curves are the evolution of the complexities with the two reheating cases considered. 
In both cases, the mode reenters the horizon after reheating; consequently, the imprint of reheating is identical in both cases (i.e., same for all $\wre<1/3$). 
The figures in the lower panel are for the spectral index $n_s=0.964$ with $N_k=55.05$. 
The reheating $e$-folds are now extended to $\nre=(9.71,38.83)$. 
Here, for reheating with $\wre=0.25$, the mode reenters the horizon during reheating, while it reenters in the RD for the other case. 
Correspondingly, the evolution of the complexity measures is different in this case. 
Similar patterns can be found when $\wre>1/3$. 
In general, the growth of OTOC and quantum discord is proportional to the number of e-folds when the modes remain outside the horizon, i.e., $\ln(|\mathcal{C}_k^T|^2)\sim S(\hat{\rho}_k)\sim 2N_{\rm out}$.
When the modes reenter the horizon, the amplitude of both the complexity measures \textit{freezes-in} to constant values depending on the expansion history.
\par
Interestingly, the evolution of circuit complexity and its slope, although dependent on the details of the reheating epoch, the final value of these quantities after horizon reentry follows the same pattern as the above two complexity measures. It should not come as a surprise as the final value depends on the value at the horizon reentry and the EOS of the phase. Thus, for modes that reenter during RD, although they have different evolution during reheating, the evolution after reentry follows the same pattern as the original perturbation modes, which is the same across all the complexity measures that we have checked. As the complexity measures in Fig (\ref{fig:evolve_ccomp2}) are highly oscillatory after mode reentry, for convenience, we have again shown the evolution of the slope of the complexity after reentry by taking the oscillation average in Fig (\ref{fig:evolve_ccomp3}).
\par
In summary, we have studied the evolution of different complexity measures for the primordial curvature perturbation, paying particular attention to the effects of the reheating phase. 
Macroscopically, reheating is a phase between inflation and RD Universe defined by an effective fluid with an EOS parameter.  
For a given inflationary potential and specifying this EOS for reheating, the number of reheating $e$-folds is required to specify the evolution fully. 
In~\ref{sec:reh_cons}, we described the reheating constraints analysis where we determined the reheating $e$-folds as a function of the scalar spectral index.
Here, we have found a very general characteristic for the evolution of cosmological perturbations when we take the reheating constraints consistently. 
We showed that reheating constraints, in addition to determining the duration of reheating, also classifies the evolution of perturbations into three class depending on the reheating EOS: (i) $\wre<1/3$, (ii) $\wre=1/2$, and (iii) $\wre>1/3$. We found that if a mode reenters the horizon after reheating, the evolution of various correlations depends on the above three classes and not on the individual EOS within the class. 
For the canonical reheating scenarios with the allowed range of the scalar spectral index from Planck~\cite{Planck:2018jri}, the evolution of the correlations for the large-scale modes observables on the CMB scales will also be categorized into such classes. 
For small-scale modes that enter during reheating, the correlation will have signatures of each reheating EOS separately. 
This, in turn, implies that studying various complexity measures and correlations of primordial perturbations on the observable scales probed by CMB-based experiments can only give us limited information about the reheating phase. We can, at best, infer the class to which they belong, not the separate EOS within the class.
\par
To this end, we can study various correlations for complexity measures for small-scale modes during preheating. 
The evolution of squeezing parameters during preheating is done in~\cite{Hirai:2000zj}. 
The presence of the coherently oscillating background inflaton, the growth of inhomogeneities, and eventually the backreaction effects will require treating the system as interacting quantum fields~\cite{Amico:2007ag,Song:2011gv,Borgonovi:2016qct,Bhattacharyya:2018bbv,Ando:2020kdz,Choudhury:2022btc,Choudhury:2022xip}. 
In the same vein, another important aspect at the end of preheating---in this context---will be the study of their thermalization properties~\cite{Luitz:2017sqb,Mueller:2021gxd}).
We also defer the study of the microscopic details of the signature of chaos from the complexity measures for a future study. 
Finally, we conclude by noting that it has been pointed out that a single OTOC fails to capture chaotic behavior for single-particle quantum chaotic systems, such as the well-known stadium billiards model~\cite{Hashimoto:2017oit,Rozenbaum:2016mmv,Rozenbaum:2018sns}, or chaotic lattice systems, such as spin chains~\cite{Gharibyan:2018fax}. 
Indeed, as pointed out in~\cite{Bhattacharyya:2019txx,Haque:2020pmp}, the out-of-time-order correlation is not the single means of probing quantum chaos within a system. 
A suite of tools for diagnosing quantum chaos, as described in references such as~\cite{Kudler-Flam:2019kxq,Bhattacharyya:2019txx,Haque:2020pmp}, encompassing various methods, including the OTOC, quantum circuit complexity.
For instance, one such effective extension of the OTOC is made by generalizing the classical matrix of phase space deviations is the quantum Lyapunov spectrum~\cite{Gharibyan:2018fax}.
These comprehensive toolkit offers a diverse perspectives on the study of quantum chaos and will be worth looking in detail for the future.
It will also be worth exploring these measures with the signatures and effects of the reheating and preheating phase. 

\medskip
\section*{Acknowledgments}
P. S. thanks Jondalar Ku\ss{} for useful suggestions and Rathul Nath Raveendran for helpful discussion. 
This study was financially supported by Seoul National University of Science and Technology.
\providecommand{\href}[2]{#2}\begingroup\raggedright\endgroup

\end{document}